\documentclass[12pt]{article}

\usepackage{amsmath,amssymb,amsthm,hyperref,bm}
\usepackage{authblk}
\usepackage{graphicx}
\usepackage{extarrows} 
\usepackage{xcolor}
\usepackage{subfig}
\usepackage{hyperref}

\definecolor{darkgreen}{rgb}{0.0, 0.5, 0.0}

\newcommand{\ii}{\mathrm{i}}
\newcommand{\N}{\mathcal{N}}
\newcommand{\C}{\mathcal{C}}
\newcommand{\Vol}{\mathcal{V}}
\newcommand{\A}{\mathcal{A}}
\newcommand{\ta}{\tilde{\bm{a}}}
\renewcommand{\j}{\iota}
\newcommand{\vx}{{\vec{x}}}

\begin{document}
\begin{titlepage}
\renewcommand{\thepage}{ }

\title{Crystal Volumes and Monopole Dynamics}

\author[*]{Sergey A. Cherkis}
\author[$\dagger$]{Rebekah Cross}
\affil[*]{\small Department of Mathematics, University of Arizona, Tucson, AZ 85721 
\tt cherkis@math.arizona.edu}
\affil[$\dagger$]{\small Department of Physics, University of Arizona, Tucson, AZ 85721
\tt rsc42@email.arizona.edu}
\date{}
\maketitle

\abstract{The low velocity dynamic of a doubly periodic monopole, also called a monopole wall or monowall for short, is described by geodesic motion on its moduli space.  This moduli space is hyperk\"ahler and non-compact.  We establish a relation between the K\"ahler potential of this moduli space and the volume  of a region in Euclidean three-space cut out by a plane arrangement associated with each monowall.  
}

\end{titlepage}

  \newpage
  \setcounter{tocdepth}{2}
  \tableofcontents

  \newpage
\section{Introduction}
A monopole wall or a {\em monowall} is a BPS monopole on $\mathbb{R}\times S^1\times S^1.$  The latter space is endowed with coordinates $(x,\theta,\varphi)$ and the product Euclidean metric $dx^2+d\theta^2+d\varphi^2$  with respective circle radii  $r_\theta$ and $r_\varphi$, i.e. $\theta\sim\theta+2\pi r_\theta$ and $\varphi\sim\varphi+2\pi r_\varphi.$  In detail, a monowall is a Hermitian bundle $E\rightarrow \mathbb{R}\times S^1\times S^1$ of rank $h$  with a pair $(A,\Phi)$ consisting of a connection $A$ on $E$ and a Higgs field $\Phi$, which is  an endomorphism of $E,$ satisfying the Bogomolny equation
\begin{align}\label{BogEq}
*D_A\Phi=-F_A.
\end{align}
Here $F_A$ is the curvature of the connection (so in a local trivialization the curvature two-form is $F_A=dA+A\wedge A$ where $A$ is the connection one-form), $*$ is the Hodge star operator, and $D_A$ the covariant differential. 

We impose the same asymptotic condition as in \cite{Cherkis:2012qs}, namely that the eigenvalues of the Higgs field grow at most linearly, having the form 
\begin{align}\label{Eq:BC}
\Phi&=\frac{\ii}{2\pi r_\theta r_\varphi} \mathrm{diag} (Q^\j_\pm x + M^\j_\pm) + O(|x|^{-1}),
\end{align}
as $x\rightarrow\pm\infty,$ 
and the connection one-form has the form 
$$\A=\frac{-\ii}{2\pi r_\theta r_\varphi} \mathrm{diag}\left(Q^\j_\pm\frac{\theta d\varphi -\varphi d\theta}{2} + r_\varphi\chi^{\theta,\j}_\pm d\theta + r_\theta\chi^{\varphi,\j}_\pm d\varphi \right)+O(|x|^{-1}).$$   
Here $\j=1,2,\ldots,h.$ 
See \cite[Sec.4]{Cherkis:2012qs} for the detailed discussion of charges $Q^\j_\pm\in\mathbb{Q}$, consistency conditions, and field asymptotics.

As argued in \cite{Cherkis:2012qs}, it is natural to enlarge the scope of our problem by allowing for Dirac-type monopole singularities at some points $p_1^+, \ldots, p_{v_+}^+$ and $p_1^-, \ldots , p_{v_-}^-$ in $\mathbb{R}\times S^1\times S^1.$  At these points the Higgs field  has (up to unitary gauge transformation) a prescribed singularity 
\begin{align}\label{Eq:DiracSing}
\Phi=\ii
\begin{pmatrix}
\frac{\pm1}{2|\vec{x}-\vec{p}_\sigma^\pm|} &0_{1\times(n-1)}\\0_{(n-1)\times 1}&0_{(n-1)\times(n-1)}
\end{pmatrix}
+O(|\vec{x}-\vec{p}_\sigma^\pm|^0).
\end{align}

The first study of monopole walls, that we are aware of, was undertaken by Ki-Myeong Lee in \cite{Lee:1998isa}, where the deformation theory of monopole walls with arbitrary compact simple Lie gauge group was studied. 
The numerical study by Richard Ward of an $SU(2)$ monowall appeared  in \cite{Ward:2006wt} and \cite{Ward:2011zz}.  The spectral curve was used in \cite{Cherkis:2012qs} to study the deformation theory of $U(h)$ monowalls. Hamanaka et al  \cite{Hamanaka:2013lza} used monowall scattering  to compute the moduli space asymptotic metric for $U(2)$ monowalls.  The interior of these moduli spaces was probed by Maldonado and Ward in \cite{Maldonado:2014gua} via special geodesics. A systematic description of the asymptotic region of the monowall moduli space and  classification of the monowall moduli spaces of real dimension four appeared in \cite{Cherkis:2014vfa}.  For a general $U(N)$ monopole, the asymptotic moduli space metric in the regime of widely separated constituents was found in \cite{Cross:2015hla}. 
Monowalls relate to a number of significant problems involving  
non-abelian Hodge theory \cite{2019arXiv190203551M}, 
mirror symmetry \cite{2018arXiv181005985T}, 
Calabi-Yau moduli spaces and quantum gauge theories in five dimensions \cite{Cherkis:2014vfa,Closset:2018bjz},
and
integrable systems \cite{Sciarappa:2017hds}.

\subsection{Spectral Data of a Monowall}\label{Sec:Spectral}
The Bogomolny equation \eqref{BogEq} implies the compatibility of the following linear system
\begin{align}\label{Eq:LinSys}
\left\{
\begin{array}{l}
(D_{\varphi}+ \ii \Phi) V(x,\theta,\varphi)=0,\\
(D_x+\ii D_{\theta}) V(x,\theta,\varphi)=0.
\end{array}
\right.
\end{align}
Here $D_x=D_{\frac{\partial}{\partial x}}$ is the covariant derivative along the $x$-direction, etc. 
It follows that the holonomy $W(s):=V(x,\theta,2\pi r_\varphi)V(x,\theta,0)^{-1}\in GL(h,\mathbb{C})$ around the $\varphi$-direction has eigenvalues that are meromorphic in the complex coordinate 
$$s:=\exp\frac{x+\ii\theta}{r_\theta}\in\mathbb{C}^*.$$ 
 This motivates introducing  the holomorphic spectral  curve 
 $$\mathbb{S}_\varphi=\left\{(s,t)\, |\, \mathrm{det}\,(t-W(s))=0\right\}\subset\mathbb{C}^*\times\mathbb{C}^*.$$ 
Moreover, the asymptotic conditions \eqref{Eq:BC} and prescribed Dirac singularity conditions \eqref{Eq:DiracSing}  ensure that this spectral curve is algebraic, given by $P(s,t)=0$, with the spectral polynomial 
\begin{align}
P(s,t)=Q(s)\,\mathrm{det}\big(t-W(s)\big)=\sum_{(m,n)\in\N} C_{m,n} s^m t^n.
\end{align}
Here $Q(s)$ is the lowest degree common multiple of the denominators of the  rational functions $q_j(s)$ appearing as coefficients of the characteristic polynomial $\mathrm{det}(t-W(s))=t^n+q_1(s)t^{n-1}+q_2(s)t^{n-2}+\ldots+q_n(s).$  This defines $P(s,t)$ up to an overall constant nonzero factor.  This ambiguity can be fixed, if desired, by imposing the dictionary order on the vertices $(m,n)\in\N$ and requiring the coefficient $C_{m_0,n_0}$ for the minimal vertex $(m_0,n_0)$ to be one. 
The Newton polygon $\N$  is the minimal convex hull of all the points $(m,n)\in\mathbb{Z}\times\mathbb{Z}$ for which $C_{m,n}\neq 0$.  The height of $\N$  is equal to the monopole bundle rank $h.$

Note that our preferential treatment of the $\varphi$ coordinate leading to the definition of the spectral curve was somewhat arbitrary.  One can instead consider the modified holonomy around the $\theta$ direction and obtain a different spectral curve $\mathbb{S}_\theta,$ now covering the $\mathbb{C}^*$ factor with coordinate $s'=\exp\frac{x-\ii\varphi}{r_{\varphi}}$.

Let Per$(\N)$ denote the set of integer perimeter points of $\N$ and let Int$(\N)$ denote the set of its integer interior points.  As demonstrated in \cite{Cherkis:2012qs}, the Newton polygon is entirely determined by the charge values $Q^\j_\pm$ and the numbers of singularities  $v_+$ and $v_-,$ while the perimeter coefficients $C_{m,n}$ with $(m,n)\in\mathrm{Per}(\N)$ are determined by the constants $M^\j_\pm\in\mathbb{R}$ and  $\chi_\pm^{\varphi,\j}\in[0,2\pi)$  appearing in the asymptotic conditions (and by the $s$-coordinates of the Dirac singularities $p_\sigma^{x,\pm}+\ii p_\sigma^{\theta,\pm}$). See \cite{Cherkis:2012qs} for details.  The interior coefficients, on the other hand, are some of the moduli (parameterizing the $L^2$ deformations) of the monopole solution, thus producing a family $\mathcal{B}_\N$ of curves (with fixed perimeter coefficients).  In fact, the total number of real moduli of a monowall is equal to four times the number of internal points: $4\times|\mathrm{Int}(\N)|$ and the moduli space is the universal Jacobian fibration of this family $\mathcal{B}_\N.$

We shall focus on the region in the moduli space with large $C_{m,n}$ and large differences between them (as specified in Section~\ref{Sec:Subwalls} using the secondary fan).  The generic curve  $\mathbb{S}_\varphi$ for a family $\mathcal{B}_\N$ is a punctured Riemann surface of genus $|\mathrm{Int}(\N)|$  with $|\mathrm{Per}(\N)|$ punctures.  Since, for any given monowall, $\mathbb{S}_\varphi$ is a curve of eigenvalues it (generically) comes equipped with a Hermitian eigen-line bundle $\mathcal{L}\rightarrow\mathbb{S}_\varphi$ with a flat connection $\nabla$.  The triplet $(\mathbb{S}_\varphi,\mathcal{L},\nabla)$ is the spectral data encoding the monowall solution $(A,\Phi),$ up to a gauge transformation, with its parameters and moduli correspondence as follows:
\begin{itemize}
\item
The holonomy of $\nabla$  around each puncture is valued in $U(1)$ and is determined by the asymptotic conditions.  This is how the $|\mathrm{Per}(\N)|$ triplets of parameters $(M^\j_\pm,\chi_\pm^{\theta,\j},\chi_\pm^{\varphi,\j})$ of the boundary conditions translate to the spectral data \cite[Sec. 4]{Cherkis:2012qs}:   $M^\j_\pm+\ii\chi_\pm^{\theta,\j}$ determine the position of each puncture, while $\chi_\pm^{\varphi,\j}$ determines the $\nabla$ holonomy  around it. 
\item
Viewing a (generic) curve $\mathbb{S}_\varphi$ as a sphere with $|\mathrm{Int}(\N)|$ handles, one can associate each handle to an internal point $(m,n)$ of $\N$ and choose a symplectic homology basis 
$\{A_f,B_{f'} | A_f\cap B_{f'}=\delta_{ff'}, A_f\cap A_{f'}=0=B_f\cap B_{f'}\}$ 
of the compactified Riemann surface $\overline{\mathbb{S}}_\varphi$ with each pair $(A_f,B_f)=(A_{m,n},B_{m,n})$ associated to the $f=(m,n)$-th handle.  Thus, each internal point $f=(m,n)\in\N$ has four moduli associated to it: two real moduli $R_f$ and $\Theta_f$ in $C_{m,n}=\exp\frac{R_f+\ii \Theta_f}{r_\theta}$ and two moduli $\Phi_f\sim \Phi_f+2\pi r_\varphi$ and $T_f\sim T_f+2\pi$ specifying the holonomies $e^{\ii\frac{\Phi_f}{r_\varphi}}$ and $e^{\ii T_f}$ of $\nabla$ around the cycles $A_f$ and $B_f,$  respectively.
\end{itemize}

Let us emphasize an important distinction between {\em parameters} and {\em moduli}.  Variations of moduli correspond to $L^2$ deformations of the solution $(A,\Phi)$ of the Bogomolny equation \eqref{BogEq}, while variations of parameters result in deformations of the solution that are not square integrable.  Physically, moduli correspond to all directions in the space of (gauge equivalence classes of) solutions  that have finite mass, while the parameters are the remaining transverse coordinates.  As a result, a monowall can slowly evolve in time with  moduli changing, while all parameters will have to remain fixed, since their time evolution would require infinite energy.  In other words the space of all monowalls with the given Newton polygon $\N$ is fibered over the parameter space.  The base is parameterized by the $3|\mathrm{Per}(\N)|$ parameters and the fiber is what we call the moduli space. The coordinates on the moduli space are the $4|\mathrm{Int}(\N)|$ moduli.  The $L^2$ norm on the tangent space of pairs $(A,\Phi)$ induces the metric on each moduli space.

\subsection{The Crystal}
Given a monowall and its spectral polynomial with coefficients $C_{m,n}$, set $R_{m,n}:=r_\theta \ln |C_{m,n}|$ and consider the set of planes 
\begin{align}\label{Planes}
\{(x,y,z)\, |\, z=m x + n y + R_{m,n}\}\subset\mathbb{R}^3.
\end{align}
Let us call the convex domain above all of these planes the {\em cut crystal}: 
\begin{align}
\C_{cut}=\{(x,y,z)\, |\, z\geq m x + n y + R_{m,n},\ \forall (m,n)\in\N\}.
\end{align}
Its surface is the graph of the function 
\begin{align}\label{Eq:Max}
M(x,y)=\max_{(m,n)\in\N}  \{ m x + n y + R_{m,n} \}.
\end{align}
The shape of the cut crystal depends on the moduli (and parameters) and we shall be interested in how its volume changes with the change in the moduli $R_{m,n}.$  Since the cut crystal has infinite volume, to keep track of these changes, let us also consider the domain above all of the perimeter planes only:
\begin{align}
\C_{0}=\{(x,y,z)\, |\, z\geq m x + n y +R_{m,n},\ \forall (m,n)\in\mathrm{Per}(\N)\}.
\end{align}
Call it the  the {\em blocked crystal}.  Its surface is the graph of the function 
$$m(x,y)=\max_{(m,n)\in\mathrm{Per}(\N)}  \{ m x + n y + R_{m,n} \}.$$ 
It is completely determined by the asymptotic conditions and is independent of the moduli, and it satisfies $m(x,y)\leq M(x,y).$  
Thus, clearly, $\C_{cut}\subseteq \C_{0}$ and the planes corresponding to the interior points of $\N$ cut $\C_{cut}$ out of the blocked crystal $\C_{0}$.  

We call the volume of the difference of the two crystals $\C_0$ and $\C_{cut}$ the {\em cut volume} 
\begin{align}\label{Eq:CutV}
\Vol(R_f):=\mathrm{Vol}\,(\C_0\setminus \C_{cut})=\mathrm{Vol}\{(x,y,z)\, |\, m(x,y)\leq z\leq M(x,y)\}.
\end{align} 
It is a function of $|\N|$ variables $R_f$, one for each integer point of $\N.$ 

Intuitively, for large moduli a monowall would split into subwalls, as demonstrated in \cite{RThesis}.  As argued in Section~\ref{Sec:Subwalls}, the subwall positions are well approximated by the $x-$positions of the vertices of this cut crystal. 
It was conjectured in \cite{Cherkis:2014vfa} that the K\"ahler potential of a monowall moduli space is related to this cut volume \eqref{Eq:CutV}.  This paper refines this conjecture and proves it.  This is based on the asymptotic metric found in \cite{Cross:2015hla}, obtained by analyzing subwall dynamic interactions via the Gibbons-Manton approach \cite{Gibbons:1995yw},  reviewed in Sections~\ref{Sec:Interaction}.  This metric approximates the metric on the moduli space end with exponential accuracy.  The  K\"ahler potential of this asymptotic metric is presented in Section~\ref{Sec:Kahler}.  This K\"ahler potential, in turn, is the Generalized Legendre Transform (GLT) of Lindstr\"om and Ro\v{c}ek \cite{Lindstrom:1987ks,Hitchin:1986ea} of the function $G.$
The main result of this paper is that the GLT function $G$ encoding the asymptotic monowall metric  equals the cut volume:
\begin{align}
G=\Vol.
\end{align}
The exact meaning and the proof of this relation are spelled out in Section~\ref{Sec:Relation}.  
It can be summarized as follows:\par 
in the regime of far separated subwalls {\em the monowall  K\"ahler potential}  {\em is the Generalized Legendre Transform of the cut volume.}

\section{Subwall Positions and  Spectral Curve Branch Points}\label{Sec:Subwalls}
As monowall moduli increase, the monowall splits into subwalls.  Let us explore the dependence of these subwall positions on the moduli.
\subsection{The Secondary Fan and the Monowall Spine}
\begin{figure}[h!]
\centering
\subfloat[]
{\label{}\includegraphics[width=0.2\textwidth]{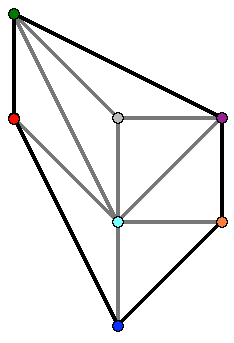}}
\hspace{0.2\textwidth}
\subfloat[]
{\label{}\includegraphics[width=0.2\textwidth]{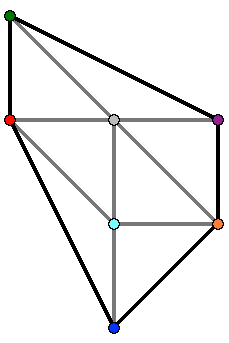}}

\subfloat[]
{\label{}\includegraphics[width=0.35\textwidth]{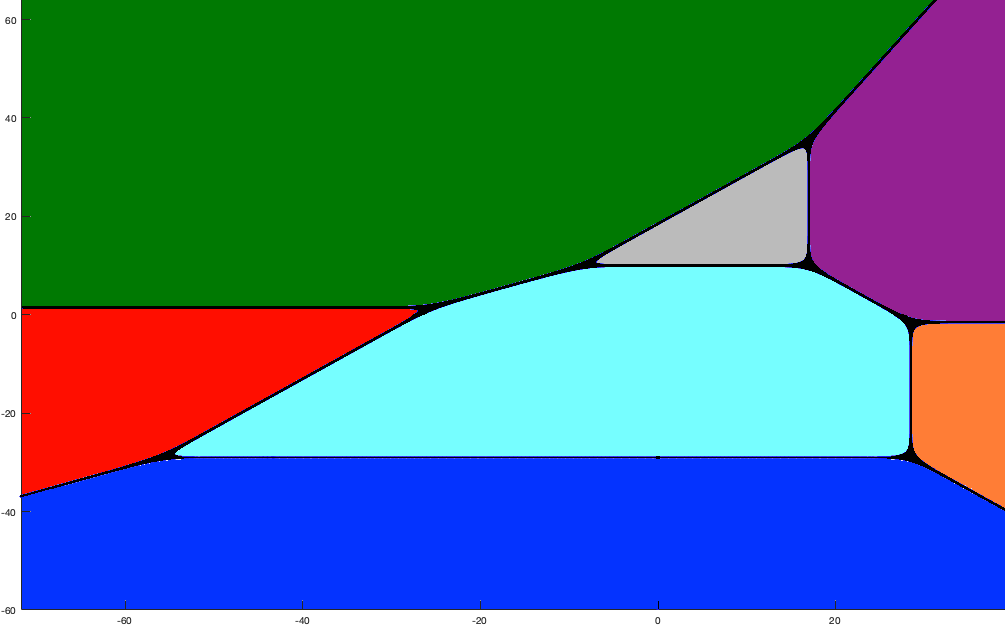}}
\qquad
\subfloat[]
{\label{}\includegraphics[width=0.35\textwidth]{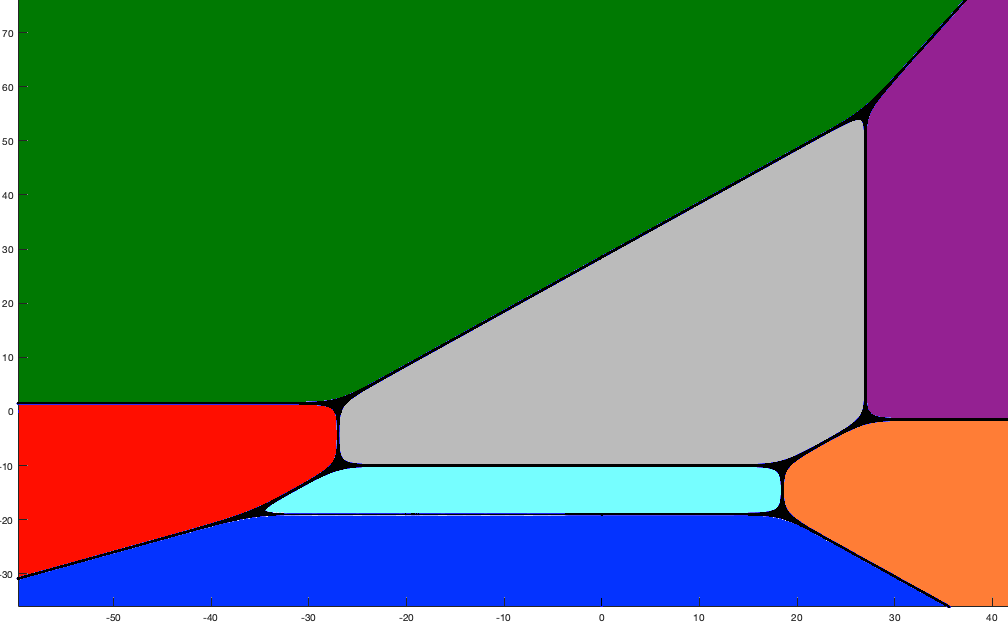}}
\caption{Newton polygon $\N$ with colored integer points and  two examples of its regular triangulations (a) and (b). 
The corresponding spines in black and their color-coded faces (c) and (d), with each face corresponding to an integer point of $\N$ in (a) and (b), respectively.}
\label{Fig:Correspondence}
\end{figure}

There is significant information about the monowall contained in the cut crystal.  Its surface consists of 
\begin{enumerate}
\item
faces (each face contained in one of the planes \eqref{Planes} and thus each has an associated integer point  $f=(m,n)\in\N$), 
\item 
edges at which these faces meet, and  
\item
vertices.
\end{enumerate}  
The projection of the cut crystal edges and vertices on the $(x,y)-$plane is a graph, that we call the {\em spine}, as illustrated in Figure~\ref{Fig:Correspondence}.  From this description the spine is  dual to a {\em regular subdivision} \cite[Chs. 6 and 7]{GKZ} of the Newton polygon $\N,$ in which the two integer points of $\N$ are connected by an edge if and only if the corresponding faces of the cut crystal meet at a crystal edge.   Each spine edge is normal to the correspoinding edge of the subdivision of $\N.$

\begin{figure}[h!]
\centering
\subfloat[]
{\label{}\includegraphics[width=0.3\textwidth]{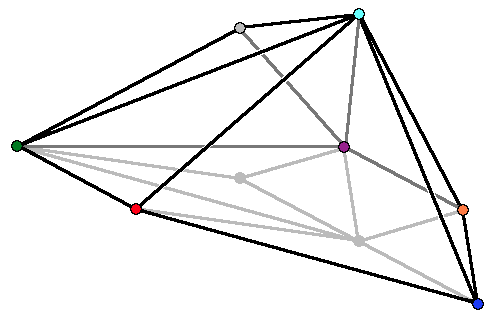}}
\hspace{0.1\textwidth}
\subfloat[]
{\label{}\includegraphics[width=0.3\textwidth]{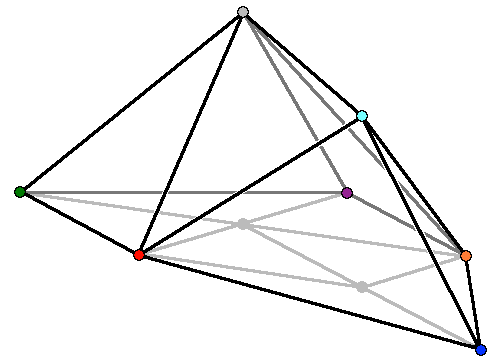}}

\subfloat[]
{\label{}\includegraphics[width=0.35\textwidth]{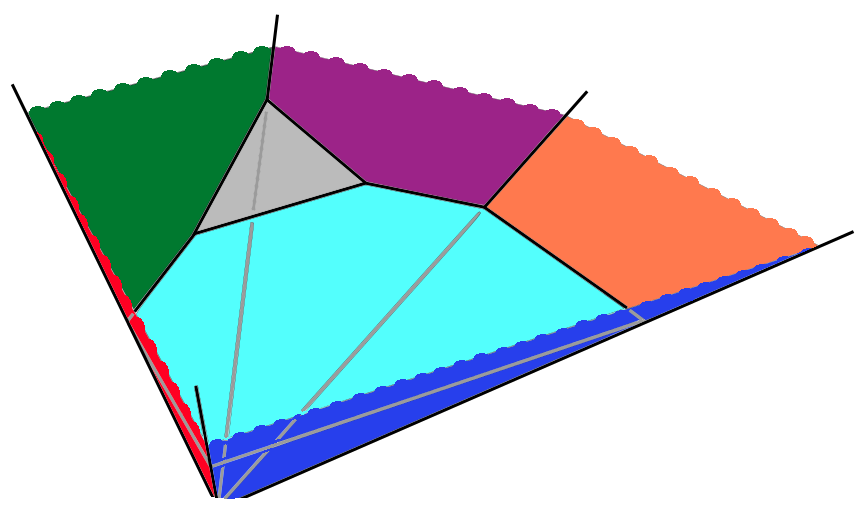}}
\qquad
\subfloat[]
{\label{}\includegraphics[width=0.35\textwidth]{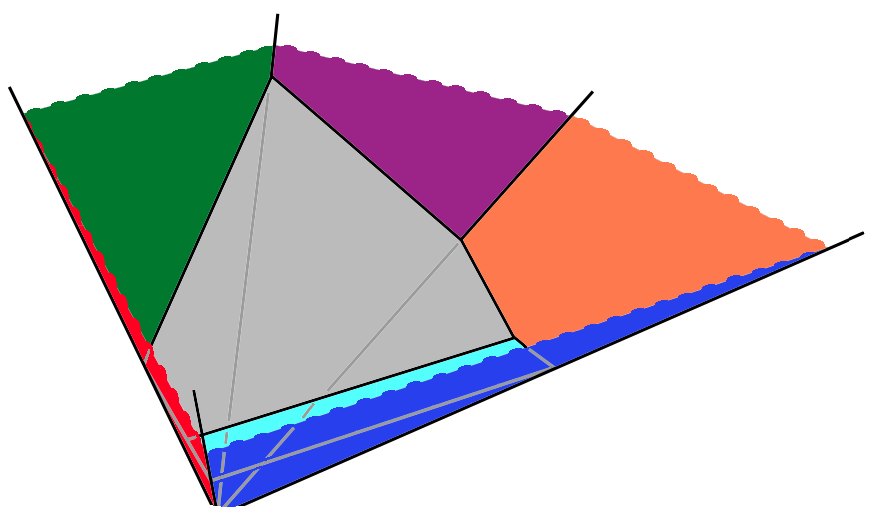}}
\caption{Two examples of the tent functions (a) and (b) and their corresponding plane arrangements (c) and (d) for the  Newton polynomial 
$F(s,t)= \textcolor{blue}{5 s} + \textcolor{cyan}{A s t} - \textcolor{orange}{5 s^2 t} + \textcolor{red}{20 t^2} + \textcolor{gray}{B s t^2} + \textcolor{violet}{20 s^2 t^2} - \textcolor{darkgreen}{5 t^3}$ with  
$(A,B)=(120,27)$ (left) and $(A,B)=(20,90)$ (right).}
\label{Fig:Correspondence2}
\end{figure}
A regular subdivision is defined in the following way. Consider a real valued function $l(m,n)$ on the integer points of $\N$ and the convex hull in $\mathbb{R}^3$ of the set of downward rays $\{ (m,n,z)\, |\,  (m,n)\in\N, z\leq l(m,n)\}.$  The part of this hull's surface that is not vertical is a graph of a concave  function over the interior of $\N$ in $\mathbb{R}^2.$ Let us call it the {\em tent function}. It is piecewise linear, with corners at (some of the points) $(m,n,l(m,n))$. The edges of this surface project onto $\N$ giving a {\em regular subdivision} of $\N.$ (Note, in some of the literature, e.g. in \cite{GKZ} itself, such subdivisions  are called  {\em coherent subdivisions} instead of {\em regular subdivisions}.)  In our case, choosing $l(m,n)=R_{m,n}=r_\theta \ln|C_{m,n}|$ results in a subdivision dual to the monowall spine.  Also, note that the resulting tent function is the negative of the Legendre transform of the cut crystal surface function $M(x,y)$ of Eq.\eqref{Eq:Max}.

As a result, the space $\mathbb{R}^{|\N|}$ with coordinates $R_f=R_{m,n}$ is subdivided into cones labelled by regular subdivisions of $\N.$  These cones form the {\em secondary fan} $F(\N)$ of $\N.$  Moving to infinity within a given cone results in the monowall splitting into subwalls of certain types, determined by the elements of that corresponding subdivision. Each polygon appearing in the subdivision corresponds to a subwall.  

The secondary fan $F(\N)$ is encoded in the secondary polytope $\Sigma(\N).$  In fact, the two are dual to each other: each ray of $F(\N)$ is normal to a face of $\Sigma(\N)$ and  two rays are connected by a wedge if the corresponding faces of $\Sigma(\N)$ share an edge. The $i-$th coordinate of the vertex of $\Sigma(N)$ can be read off from its corresponding regular triangulation as the total area of the triangles for which the $i-$th integer point of $\N$ is a vertex.  See \cite{GKZ} for many fascinating details.

There is a partial order on all regular subdivisions given by refinement.  The maximally refined subdivisions are the regular triangulations with each triangle of area\label{FT:AreaNorm}\footnote{As in \cite{GKZ}, we normalize the area of a basic triangle with vertices $(0,0),(1,0),$ and $(0,1)$ to be one, instead of a half.}  $1$.  
This is the case we are most interested in here, as it corresponds to the monowall maximally split into elementary subwalls.  

Each regular triangulation labels a cone (of maximal dimension) in the secondary fan (see \cite[Ch.6]{GKZ}), with other regular subdivisions labelling its lower-dimensional cones. According to \cite[Sec.5]{Cherkis:2014vfa}, the secondary fan is in the space $\mathbb{R}^{|\N|}$  with  coordinates $R_{m,n},$ which include both moduli and parameters. A generic direction lies in the interior of a single cone of the secondary fan and corresponds to some regular triangulation. 
Fixing all parameters gives a slice $\mathbb{R}^{|\mathrm{Int}(\N)|}$ of $\mathbb{R}^{|\N|}=\mathbb{R}^{|\mathrm{Per}(\N)|}\times \mathbb{R}^{|\mathrm{Int}(\N)|}$. The intersection of this slice with the secondary fan divides this slice into regions, some compact and some noncompact.

The `down-facing' cones of the secondary fan correspond to triangulations not involving any internal points of $\N$ as triangle vertices. (These form the associahedral face of the secondary polytope, its largest face.)
The maximally refined triangulations described above correspond to `upward-facing' cones. It is these latter that correspond to  asymptotic directions in the monowall moduli space (the noncompact regions of the secondary cone subdivision of an $\mathbb{R}^{|\mathrm{Int}(\N)|}$ slice).  Such, regular triangulations describe generic asymptotics  of a monowall moduli space.

To summarize, for a regular triangulation there is the following correspondence illustrated in Figure~\ref{Fig:Correspondence}:
\begin{enumerate}
\item
each face $f$ of the spine\footnote{A spine face is the projection of a  face of the cut crystal.} corresponds to an integer point $(m,n)$ in the Newton polygon $\N,$
\item
each edge of a spine is an interface between faces $f_1$ and $f_2$ and it is orthogonal to the edge of the triangulation connecting the two corresponding integer points $(m_1,n_1)$ and $(m_2,n_2)$ of $\N,$ and
\item
each vertex $a$ of the spine corresponds to a triangle $\Delta_a$ of the triangulation Triang$(\N)$ of $\N.$
\end{enumerate}
Clearly, point 2. above can be stated as:  the spine edge connecting vertices $a$ and $b$ is orthogonal to an edge of the triangulation of the Newton polygon that is shared by the triangles $\Delta_a$ and $\Delta_b.$

\subsection{Subwall Positions}
Let us now explore the generic asymptotic region in the monowall moduli space by fixing a regular maximal triangulation and moving along a ray in the corresponding upward facing cone of the secondary fan.  Each triangle of this triangulation corresponds to a vertex of the spine.  We claim that (up to a constant, moduli independent, shift) the $x-$position of this vertex is the position of the corresponding subwall into which the monowall splits.  
To be exact, we understand the position of the subwall to be the point of (partial) gauge symmetry restoration, i.e. the branch point of the spectral curve $\mathbb{S}_\varphi.$

\subsubsection{Spine Vertex}\label{Sec:SpineV}
Consider a triangle $\Delta_a$ of the regular maximal triangulation. Say its vertices are $(m_1,n_1),(m_2,n_2),$ and $(m_3,n_3)$ ordered counterclockwise.  
A crystal vertex corresponding to that triangle  is positioned at the point $(x_a,y_a,z_a)$ satisfying
\begin{align}
m_1 x_a+n_1 y_a+R_{m_1,n_1}&=z_a,\nonumber\\
\label{Eq:Planes}
m_2 x_a+n_2 y_a+R_{m_2,n_2}&=z_a,\\
m_3 x_a+n_3 y_a+R_{m_3,n_3}&=z_a,\nonumber
\end{align}
with solution
\begin{align}\label{Eq:den}
\left(\begin{smallmatrix}
x_a\\y_a\\z_a
\end{smallmatrix}\right)
=\frac{-1}{
\left|\begin{smallmatrix}
m_1-m_3 & m_2-m_3\\
n_1-n_3 & n_2-n_3
\end{smallmatrix}\right|
}
\left(\begin{smallmatrix}
n_3-n_2&n_1-n_3&n_2-n_1\\
m_2-m_3&m_3-m_1&m_1-m_2\\
m_2n_3-m_3n_2&m_3n_1-m_1n_3&m_1n_2-m_2n_1
\end{smallmatrix}\right)
\left(\begin{smallmatrix}
R_{m_1,n_1}\\R_{m_2,n_2}\\R_{m_3,n_3}
\end{smallmatrix}\right).
\end{align}
Since the triangulation is maximal and the points $(m_i,n_i)$ are numbered counterclockwise, the triangle has minimal area, thus the denominator in \eqref{Eq:den} is $+ 1$ and the $x-$position of the spine vertex is $x_a=(n_2-n_3)R_{m_1,n_1}+(n_3-n_1) R_{m_2,n_2}+(n_1-n_2)R_{m_3,n_3}.$  To simplify the notation, let $R_j=R_{m_j,n_j}$ and $\delta n_{ij}=n_i-n_j$ and same for other quantities, then
\begin{align}\label{Eq:SpineVer}
 x_a&=\delta n_{23}R_1+\delta n_{31}R_2+\delta n_{12}R_3\\
 &=-n_1\delta R_{23}-n_2\delta R_{31}-n_3\delta R_{12}\\
 &=\delta n_{23}\delta R_{12}-\delta n_{12}\delta R_{23}.
\end{align}

\subsubsection{Spectral Curve Branch Points}
The holonomy of $D_\varphi+\ii\Phi$ breaks the $U(n)$ gauge symmetry, and when the gauge symmetry is maximally broken to $U(1)^h$ the Bogomolny equation for the resulting $U(1)^h$ fields is Abelian, implying that the $U(1)^h$  Higgs field is harmonic. Thus, at large distances the Higgs field is linear. Therefore, it is exactly the regions where the gauge symmetry is at least partially restored that can be viewed as the sources of electromagnetic fields.  This argument (at least in the limit of large separation of all subwalls) associates the magnetic charge to the regions where some eigenvalues of the holonomy coincide. In other words, the monowall consists of subwalls positioned at the branch points of the spectral curve. These subwalls carry magnetic $U(1)^h$ charges and the magnetic field is constant between them.

Our immediate task is finding the locations of the branch points, in particular, their $x-$coordinates, with $x=r_\theta\ln|s|,$  and comparing them with the $x-$locations of the spine vertices found above. As moduli become large, so does the spectral curve.  To keep the whole curve in view, we rescale the coordinates accordingly. To begin, let $s=\exp(\frac{1}{\hbar}\frac{a}{r_\theta}+\ii\frac{\theta}{r_\theta}), t=\exp(\frac{1}{\hbar}\frac{b}{r_\theta}+\ii\alpha),$ and 
$C_{m,n}=\exp(\frac{1}{\hbar}\frac{l_{mn}}{r_\theta}+\ii\frac{\Theta_{mn}}{r_\theta}).$    We consider the relevant locations $a$ of the branch points as $\hbar$ is sent to zero, which corresponds to the large moduli region.

From the basic Puiseux expansion, each branch point is governed by three relevant monomials of the spectral polynomial $P(s,t)=\sum_{(m,n)\in\N}C_{m,n}s^m t^n$ (see \cite{RThesis}   for details), corresponding to the vertices of some triangle $\Delta$ of the given regular triangulation of $\N$, therefore we can focus on 
\begin{align}\label{Zoom}
C_{m_1,n_1}s^{m_1}t^{n_1}+C_{m_2,n_2}s^{m_2}t^{n_2}+C_{m_3,n_3}s^{m_3}t^{n_3}=0.
\end{align}
The other terms are exponentially  small ($\sim e^{-K/\hbar}$ with some $K>0$) in the moduli. 
If needed, relabel the vertices so that $n_3\leq n_1,n_2.$  Let $N_j=n_j-n_3$ and $M_j=m_j-m_3,$ for $j=1,2.$  If needed, exchange the indices $1$ and $2$ to have the counterclockwise orientation, so that 
$\begin{vmatrix}
N_1&N_2\\ M_1&M_2
\end{vmatrix}=1.$  Now, the above equation reads 
$C_{m_1,n_1}s^{M_1}t^{N_1}+C_{m_2,n_2}s^{M_2}t^{N_2}+C_{m_3,n_3}=0.$ In the new variables 
$S:=s^{M_1}t^{N_1}$ and $T:=s^{M_2}t^{N_2},$
Eq.\eqref{Zoom} becomes $C_{m_1,n_1} S+C_{m_2,n_2} T+C_{m_3,n_3}=0,$  thus   
$T=-\frac{C_{m_1,n_1}}{C_{m_2,n_2}}\left(S+\frac{C_{m_3,n_3}}{C_{m_1,n_1}}\right)=A(S-\alpha),$ with $A= -\frac{C_{m_1,n_1}}{C_{m_2,n_2}}$ and $\alpha=-\frac{C_{m_3,n_3}}{C_{m_1,n_1}}.$

In terms of $S$ and $T$ the original variables are
\begin{align}
\label{Eq:sS}
s&=S^{-N_2}T^{N_1}=A^{N_1}S^{-N_2}(S-\alpha)^{N_1},\\
t&=S^{M_2} T^{-M_1}=A^{N_1}S^{M_2}(S-\alpha)^{-M_1}.
\end{align}
Branching of $t$ as a function of $s$ can only occur at the branch points of $S(s),$ the solution of \eqref{Eq:sS}.  These occur at the roots of the discriminant of the polynomial 
\begin{align}
Q(S)=A^{N_1}(S-\alpha)^{N_1}-s S^{N_2}.
\end{align}
The discriminant is proportional to the resultant $R(Q,Q')$, which we now compute.

Let us list some basic properties of the resultant (see e.g. \cite{swan1962}) of a pair of polynomials:
\begin{align*}
R(f,g)&=(-1)^{\deg f\cdot \deg g} R(g,f),\\
R(gq+r,g)&=b^{\deg(gq+r) -\deg r}R(r,g),\  
\text{where\ } b \text{ is the leading coefficient of \ } g,\\
R(f_1f_2,g)&=R(f_1,g)R(f_2,g),\\
R(f,a)&=a^{\deg f}=R(a,f),\\
R(f,x-\alpha)&=f(\alpha),\\
R(x^n-\alpha, x^m-\beta)&=(-1)^m (\alpha^{m'}-\beta^{n'})^d,
\end{align*}
where $d=GCD(n,m)$, $n=n' d$ and $m=m' d.$

Clearly,
\begin{align}
Q'(S)&=N_1 A^{N_1}(S-\alpha)^{N_1-1}-N_2 s S^{N_2-1},\\
Q(S)&=A^{N_1}(S-\alpha)^{N_1}-s S^{N_2}\nonumber\\
&=\frac{S-\alpha}{N_1}Q'(S)+s \frac{N_2-N_1}{N_1} (S-\frac{N_2}{N_2-N_1}\alpha)S^{N_2-1}.
\end{align}
And the resultant is 
\begin{multline}
R(Q,Q')=(N_1 A^{N_1})^{N_1-N_2-1}R\left(s \frac{N_2-N_1}{N_1} S^{N_2-1}(S-\frac{N_2}{N_2-N_1}\alpha),Q'\right)\\
=(N_1 A^{N_1})^{N_1-N_2-1}\left(s \frac{N_2-N_1}{N_1}\right)^{N_2}
R(S,Q')^{N_2-1} 
R\left(S-\frac{N_2}{N_2-N_1}\alpha, Q'\right) \\
=(N_1 A^{N_1})^{N_1-N_2-1}\left(s \frac{N_2-N_1}{N_1}\right)^{N_2}
\left(Q'(0)\right)^{N_2-1}
Q'\left(\frac{N_2}{N_2-N_1}\alpha\right) \\
=(N_1 A^{N_1})^{N_1-N_2-1}\left(s \frac{N_2-N_1}{N_1}\right)^{N_2}
\left(  N_1 A^{N_1}(-\alpha)^{N_1-1}
\right)^{N_2-1}\\
\left(N_1 A^{N_1}\left(\frac{N_1}{N_2-N_1}\alpha\right)^{N_1-1}
-N_2 s \left(\frac{N_2}{N_2-N_1}\alpha\right)^{N_2-1}\right).
\end{multline}

It vanishes at 
\begin{align}\label{Eq:BranchAll}
s=(-1)^{N_2}\left(\frac{N_1}{N_2}\right)^{N_1}\frac{1}{(N_2-N_1)^{N_1-N_2}} \frac{C_{m_1,n_1}^{N_2}}{C_{m_2,n_2}^{N_1}}C_{m_3,n_3}^{N_1-N_2}.
\end{align}
This is the position of a branch point corresponding to the triangle $\Delta$.  
In terms of the spatial position $x=r_\theta\ln|s|=a/\hbar$ of this branch point of the spectral curve, we  have
\begin{multline}\label{Eq:Branch}
\hbar x_\Delta=a=\left((n_2-n_3) l_{m_1,n_1}+(n_3-n_1) l_{m_2,n_2}+(n_1-n_2) l_{m_3,n_3}\right)\\
+\hbar  \ln \left(\frac{N_1}{N_2}\right)^{N_1}\frac{1}{(N_2-N_1)^{N_1-N_2}},
\end{multline}
which matches the  position of the vertex of the spine \eqref{Eq:SpineVer}  up to  $O(\hbar^0)$ terms. 
The $\theta=r_\theta\mathrm{Arg} (s)$ coordinate of the branch point is read off as the imaginary part of \eqref{Eq:BranchAll}:
\begin{align}
\theta_\Delta=\left((n_2-n_3) \Theta_{m_1,n_1}+(n_3-n_1) \Theta_{m_2,n_2}+(n_1-n_2) \Theta_{m_3,n_3}\right).
\end{align}
Thus, for large values of $R_{m,n}=r_\theta\ln|C_{m,n}|=\frac{l_{m,n}}{\hbar}$ the subwalls are positioned at the $x-$locations of the vertices of the spine.  Moreover, the $x$ and $\theta$ positions of the wall associated to the spine vertex $a$  (corresponding to the triangle $\Delta_a$ of the triangulation) are expressed via the same relation
\begin{align}\label{Eq:caf}
x_a&=\sum_{f=1}^3 c_a^f R_f,&
\theta_a&=\sum_{f=1}^3 c_a^f \Theta_f,
\end{align}
where the sum is over the three spine faces containing to the vertex $a.$ When these three faces are numbered counterclockwise, the coefficients $c_a^f$ are $c_a^1= (n_2-n_3), c_a^2= (n_3-n_1),$ and $c_a^3= (n_1-n_2).$

\subsection{Subwall Charges and Inter-wall Fields}
\begin{figure}[htb]
\centering
\subfloat[][]
{\includegraphics[width=0.24\textwidth]{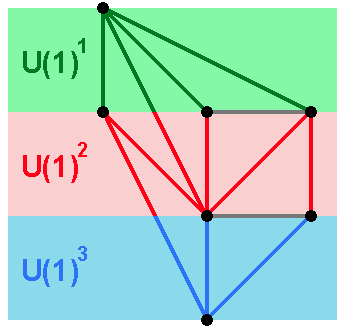}}
\hspace{0.03\textwidth}
\subfloat[][]
{\includegraphics[width=0.7\textwidth]{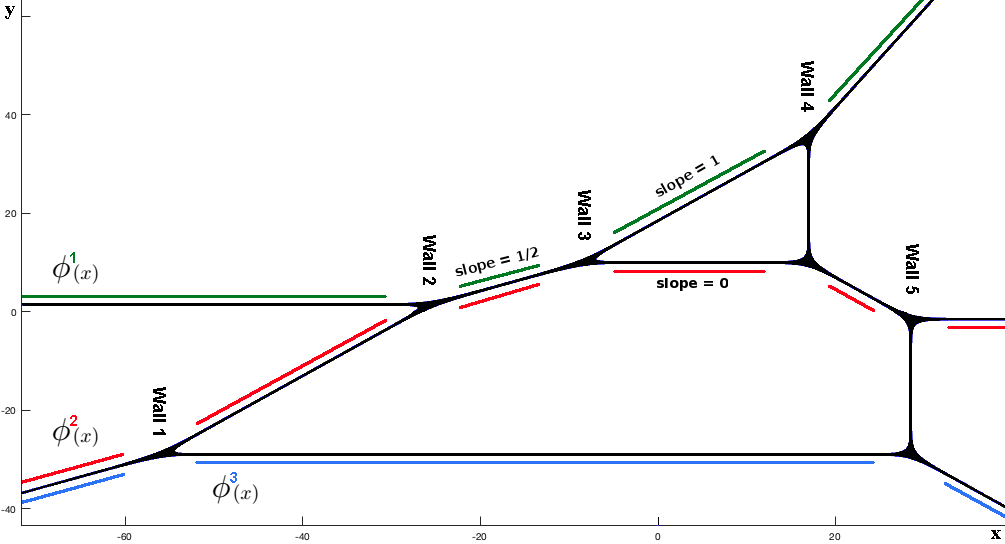}
}
\caption{Three Higgs eigenvalues of a $U(3)$ monowall, shown as green, red and blue lines over a range in $x$. \textit{Wall 3} has magnetic charges $+1/2$ and $-1/2$ in the first two (green and red) $U(1)$ factors  and charge $0$ in the third (blue) factor. Each spine edge is orthogonal to the corresponding edge of the triangulation.}
\label{Fig:Ribbons}
\end{figure}

If the vertices of the spine indicate the subwall positions, the spine edges approximate the eigenvalues of the Higgs field between the walls (to exponential accuracy in distance to the nearest wall). Away from all subwalls the $U(n)$ gauge  symmetry is broken to $U(1)^h$ with each $U(1)$ factor associated to one $D_\varphi+\ii\Phi$ holonomy eigenvalue $t^\j$.  We order these  eigenvalues in decreasing order of $y^\j=r_\theta\ln |t^\j|$ so that $y^1\geq y^2\geq \ldots\geq y^n.$  Thus, each $y^\j$ is a continuous,  piecewise linear function $\bm{\phi}^\j$  of $x$ with kinks at the spine vertices.  Away from the walls we have Higgs field
\begin{align}
\Phi=\frac{\ii}{2\pi r_\theta r_\varphi}\mathrm{diag}(y^\j)+O(e^{- \frac{d_{\mathrm{wall}}}{\Lambda}}),
\end{align}
where $d_\mathrm{wall}$ is the distance to the closest wall and the constant $\Lambda$ is the characteristic wall width,  computed in \cite[Sec.3]{Cross:2015hla}. 

Since each spine edge, orthogonal to the $(m_{ff'},n_{ff'})=(m_{f}-m_{f'},n_{f}-n_{f'})$ edge of the triangulation of $\N$, corresponds to $|n_{ff'}|$ of the ordered eigenvalues, now all non-vertical edges of the spine are labelled by the factor indices $\j$, as illustrated in Figure~\ref{Fig:Ribbons}.  If we associate each index value to some distinct color, then the spine consists of continuous lines going left to right, each piecewise linear with kinks at the spine vertices. Each colored line is a graph of a function $y^\j(x)$. It corresponds to an (approximate) Higgs diagonal  value and the slope $S_e=-\frac{n_e}{m_e}=-\frac{n_{h(e)}-n_{t(e)}}{m_{h(e)}-m_{t(e)}}$ of the line is the magnetic field of the corresponding $U(1)$ factor.  The difference $g^\j=S^\j_\mathrm{right}-S^\j_\mathrm{left}$ in line slopes   $S^\j_\mathrm{right}$ and $S^\j_\mathrm{left}$ across the wall is the magnetic charge in that $\j-$th $U(1)$ factor of the wall corresponding to this spine vertex.

Next, we interpret the resulting fields as a superposition of individual subwall contributions and explore the subwall  dynamics.

\section{Subwall Interactions}\label{Sec:Interaction}
The moduli space of a monowall is of real dimension $4|\mathrm{Int}(\N)|$ with half of the moduli being the coefficients $C_{m,n}=\exp{\frac{R_{m,n}+\ii\Theta_{m,n}}{r_\theta}},$  (for $(m,n)\in\mathrm{Int}\,\N$) of the spectral curve $\mathbb{S}_\varphi$  and the other half $(\Phi_{m,n},T_{m,n})$ parameterizing the Hermitian eigen-line bundle with a flat connection over $\mathbb{S}_\varphi.$  We view this moduli space as a three-torus fibration over the $|\mathrm{Int}(\N)|-$dimensional space $\mathbb{R}^{|\mathrm{Int}(\N)|}$ with base coordinates $R_{m,n}=r_\theta \ln|C_{m,n}|, (m,n)\in\mathrm{Int}\,\N$ and the fiber coordinates $(\Theta_{m,n},\Phi_{m,n},T_{m,n}).$ 
The space of `long' moduli and parameters $R_{m,n}$ factors as a direct product  $\mathbb{R}^{|\N|}=\mathbb{R}^{|\mathrm{Int}(\N)|}\times \mathbb{R}^{|\mathrm{Per}(\N)|}$ of the space of all `long' moduli and of the space all `long' parameters.  
As we discussed, this space $\mathbb{R}^{|\N|}$ contains the secondary fan whose maximal cones are indexed by the regular triangulations of the Newton polygon $\N.$  The space of long moduli is obtained by fixing the values of the long parameters.  This is the base space of the moduli space fibered by tori.  It traverses this fan, and the fan  subdivides it into polytopal regions, each region corresponding to a phase of the monowall.  The monowall in each phase, labelled by a triangulation Triang$(\N),$ is well  approximated (for sufficiently large  moduli) by an array of subwalls. And $a$-th subwall corresponds to a triangle $\Delta_a\in\mathrm{Triang}(\N)$ and it carries $h$ Abelian charges $(g^1_a,\ldots, g^h_a)$ (defined by the slopes of the two sides of the triangle $\Delta_a$ to which  the graph of $y_\j(x)$ is associated). Away from any subwall the  Higgs field is essentially diagonal $\Phi=\frac{\ii}{2\pi r_\theta r_\varphi}\mathrm{diag}(y^\j)+O(e^{- \frac{d_{\mathrm{wall}}}{\Lambda}})$ with
\begin{align}\label{Eq:DiagHiggs}
y^\j=\frac{Q^\j_+ +Q^\j_-}{2} x+\frac{M^\j_+ +M^\j_-}{2}
+\sum_{\Delta\in \mathrm{Triang}(\N)} \frac{g_\Delta^\j}{2} |x-x_\Delta| +O(e^{- \frac{d_{\mathrm{wall}}}{\Lambda}}).
\end{align}
Here $Q^\j_+-Q^\j_-=\sum_\Delta g^\j_\Delta$ and $M^\j_- - M^\j_+=\sum_\Delta g_\Delta^\j x_\Delta,$ and  the subwalls' positions are
$x_\Delta=(n_2-n_3)R_{m_1,n_1}+(n_3-n_1)R_{m_2,n_2}+(n_1-n_2)R_{m_3,n_3}.$

Let us discuss the meaning of \eqref{Eq:DiagHiggs} in  detail, neglecting the exponentially small terms from now on.  A single Abelian wall positioned at $(x_a,\theta_a,\varphi_a)$ produces fields $\Phi=\mathrm{diag}(\Phi^\j)$ and $A=\mathrm{diag}(A^\j)$ with
\begin{align}\label{Eq:OneWall}
\Phi^\j_a&=\frac{\ii}{2\pi r_\theta r_\varphi} \frac12 g_a^\j |x-x_a|,\\
A^\j_a&=\frac{\ii}{2\pi r_\theta r_\varphi}  \mathrm{sign}(x-x_a) \frac12g_a^\j \frac{(\varphi-\varphi_a)d\theta - (\theta-\theta_a)d\varphi}{2}.
\end{align}
The superposition of such fields has  left-right symmetric asymptotics.  To accommodate  general monowall charges, let 
\begin{align}\label{Eq:Brd}
\bar{Q}^\j&=\frac{Q^\j_+ +Q^\j_-}{2}&
&\text{and}&
\bar{M}^\j&=\frac{M^\j_+ +M^\j_-}{2},&
\end{align}
so that the total fields are
\begin{align}\label{Eq:AllStat}
\Phi^\j&= \frac{i}{2 \pi r_\theta r_\varphi} \left( \bar{Q}^\j x+ \bar{M}^\j \right)+\sum_a \Phi^\j_a, \\
A^\j&=\frac{-\ii}{2 \pi r_\theta r_\varphi} \left( \bar{Q}^\j \frac{\theta d\varphi -\varphi d\theta}{2} + r_\varphi \bar{\chi}_{\theta}^\j d\theta + r_\theta \bar{\chi}_{\varphi}^\j d\varphi \right)+\sum_a A^\j_a.
\end{align}

Now, any variation of the moduli produces some motion of the subwalls.  A moving charged wall produces Li\'enard-Wiechert potentials \cite[Eq.(16)]{Cross:2015hla}, instead of those of the static potentials of Eq.\eqref{Eq:OneWall}.  In addition, each wall has an associated electromagnetic phase.  Time dependence of this phase produces an electric charge $q_a.$  Thus, each subwall (with varying moduli) becomes a dyonic moving wall with magnetic charges $g_a^\j$, respective electric charges $q_a g_a^\j$ and velocity $\vec{V}_a.$  We spell out the explicit expressions for these potentials next. 

\subsection{Moving Dyonic Subwalls}
To avoid superficial prefactors let $\Phi^\j=\frac{\ii}{2\pi r_\theta r_\varphi}\bm{\phi}^\j$ and $A^\j=\frac{-\ii}{2\pi r_\theta r_\vartheta} \bm{a}^\j$, so that for an Abelian monowall
\begin{align}
*d\bm{a}^\j=d\bm{\phi}^\j.
\end{align}
An elementary static wall  positioned at $x=0,\theta=0$ produces 
\begin{align}
\bm{\phi}^\j(x)&=\frac12 g^\j |x|,&
\bm{a}^\j(x)&=\frac12 g^\j\, \eta_{\vx},
\end{align}
where the one-form $\eta_\vx$ satisfies $*d\eta_\vx=d|x|,$ for example 
\begin{align}
\eta_\vx=\begin{cases} \frac{\theta d\varphi- \varphi d\theta}{2},& \text{for } x>0\\ 
                                     -\frac{\theta d\varphi- \varphi d\theta}{2},& \text{for } x<0\end{cases}.
\end{align}
The superposition of such subwalls as in \eqref{Eq:AllStat} produces the functions \eqref{Eq:DiagHiggs} read off from the spine with $y^\j=\bm{\phi}^\j$.

A BPS dyon with electric charge $q$ and magnetic charge $e$   satisfies BPS equations 
$B=\frac{e}{\sqrt{e^2+q^2}}\nabla \Phi$ and $E=\frac{q}{\sqrt{e^2+q^2}}\nabla\Phi$ \cite{Coleman:1976uk}, so 
an elementary dyonic  wall can be described by the pentuple $(\bm{\phi},\bm{a}_0,\bm{a},\ta_0,\ta)$ consisting of the scalar Higgs field $\bm{\phi}$, an electro-magnetic potential consisting of the time component function $\bm{a}_0$ and a `vector' component one-form $\bm{a},$ and dual electromagnetic potentials $\ta_0$ (a function) and $\ta$ (a one-form). These are related by electromagnetic duality
\begin{align}
+d\bm{a}=B^\flat&=\tilde{E}^\flat=-d\ta_0-\dot{\ta},\\
d\bm{a}_0+\dot{\bm{a}}=-E^\flat&=\tilde{B}^\flat=*d\ta.
\end{align}
Here $E^\flat, B^\flat$ are the one-forms metric dual to the electric and magnetic vector fields, and  $\tilde{E}^\flat, \tilde{B}^\flat$ are in the same relation as the electro-magnetic dual fields $\tilde{E}:=B$ and $\tilde{B}:=-E.$ 

In these terms the fields of the $b-$th static dyonic wall with magnetic charges $g_b^\j$ and electric charges $q_b g_b^\j$ positioned at $\vec{x}=\vec{x}_b$ are
\begin{align}
\bm{\phi}^\j(x)&=\frac12  g^\j_b \sqrt{1+q_b^2} |x-x_b|,
\end{align}
\begin{align}
\bm{a}^\j(x)&=\frac12 g_b^\j \eta_{\vx-\vx_b},&
\bm{a}_0^\j(x)&=-q_b \frac12 g_b^\j |x-x_b|,\\
\ta^\j(x)&=-q_b\frac12  g_b^\j\eta_{\vx-\vx_b},&
\ta_0^\j(x)&=- \frac12 g_b^\j|x-x_b|.
\end{align}
The Lorentz boost (accompanied by the proper time delay) produced the Li\'enard-Wiechert potential produced by the moving dyonic wall.  Our focus is on the dynamics of slowly moving walls, thus, we neglect terms higher than quadratic in the resulting Lagrangian.  In particular, the typical time delay terms of the form $\sqrt{\vec{x}^2-(\vec{x} \times\vec{V})^2}$ can be safely replaced by $|x|.$  The resulting fields are 
\begin{align}
{\bm{\phi}}^\j_b(x)&= \sqrt{1+q_b^2} \frac12 g_b^\j  |x-x_b|\sqrt{1-\vec{V}_b^2}\nonumber\\
&=\frac12 g_b^\j |x-x_b|\left(1+\frac{q_b^2}{2}-\frac{\vec{V}^2}{2}\right)+o(V_b^2,q_b^2),
\end{align}
\begin{align}
{\bm{a}}^\j_b(x)&=\frac12 g_b^\j \eta_{\vx-\vx_b}- \frac12 q_b g_b^\j|x-x_b| \vec{V}_b^\flat+o(V_b^2,q_b^2),\\
{\bm{a}}_{0b}^\j(x)&=- \frac12 q_b g_b^\j |x-x_b|+\frac12 g_b^\j\eta_{\vx-\vx_b}(\vec{V}_b)+O(V_b^2,q_b^2),\\
{\ta}^\j_b(x)&=- \frac12 q_b g_b^\j \eta_{\vx-\vx_b} - \frac12 g_b^\j|x-x_b| \vec{V}_b^\flat+O(V_b^2,q_b^2),\\
{\ta}_{0b}^\j(x)&=-\frac12 g_b^\j|x-x_b|- \frac12 q_b g_b^\j\eta_{\vx-\vx_b}(\vec{V}_b)+o(V_b^2,q_b^2).
\end{align}
Here $\vec{x}=(x,\theta,\varphi),$  $\vec{V}^\flat=V^x dx+V^\theta d\theta+V^\varphi d\varphi,$ and $\eta(\vec{V})=\eta_x V^x+\eta_\theta V^\theta+\eta_\varphi V^\varphi$ is the value of the one-form $\eta$ on the vector $\vec{V}.$ From now on we drop the higher order terms in $V$ and $q.$

A dyonic wall $a$ moves in the background of fields $(\bm{\phi},\bm{a},\bm{a}_0,\ta,\ta_0)$ which are the sum of contributions of all other walls.  For example, (keeping up to quadratic terms in $V$ and $q$) the Higgs field that the $a$-th wall experiences is
\begin{align}\label{Eq:phitot}
\bm{\phi}^\j(x_a)&=\bar{Q}^\j x_a+\bar{M}^\j
+\sum_{b} {\bm{\phi}}^\j_b(x_a)\nonumber\\
&=\bar{Q}^\j x_a+\bar{M}^\j
+\sum_{b} \frac12 g_b^\j  |x_a-x_b| \left(1+\frac{q_b^2}{2}-\frac{\vec{V}_b^2}{2}\right).
\end{align}
Similarly, 
\begin{align}\label{Eq:atot}
\bm{a}^\j(x_a)&=\bar{Q}^\j \frac{\theta_a d\varphi-\varphi_a d\theta}{2}+ r_\varphi \bar{\chi}^\j_\theta d\theta
+ r_\theta \bar{\chi}^\j_\varphi d\varphi+\sum_{b} \hat{\bm{a}}^\j_b(\vec{x}_a), \\
\bm{a}^\j_0(x_a)&=\sum_{b} \hat{\bm{a}}^\j_{0b}(\vec{x}_a), \\
\ta^\j(x_a) &= \sum_{b} \hat{\ta}^\j_b(\vec{x}_a), \\
\label{Eq:adualtot}
\ta^\j_0(x_a) &= - \bar{Q}^\j x_a - \bar{M}^\j + \sum_{b} \hat{\ta}^\j_{0b}(\vec{x}_a).
\end{align}
Note, that since the fields produced by any given subwall itself vanish at its location, there are no self-interaction terms, and the sums above are extended over all walls.   
The resulting relativistic Lagrangian $\hat{L}_a$ governing the $a$-th subwall dynamics is 
\begin{align}
\hat{L}_a= \sum_\j \big\{-g_a^\j\bm{\phi}^\j(x_a) \sqrt{1+q_a^2}\sqrt{1-\vec{V}_a^2} 
&+ q_a g_a^\j [\bm{a}^\j(x_a)(\vec{V}_a) -  \bm{a}_0^\j(x_a)]\nonumber\\
&+g_a^\j[\ta^\j(x_a)(\vec{V}_a)- \ta_0^\j(x_a)]\big\},
\end{align}
with the background fields given by (\ref{Eq:phitot}--\ref{Eq:adualtot}). 
This Lagrangian $\hat{L}_a$ governing the motion of one of the subwalls should be understood as the part of the effective Lagrangian $\hat{L}$ of the whole monowall governing the motion of all subwalls.  In particular it is the part of $\hat{L}$ that contains $\vec{x}_a.$ Note, that the two subwall interaction is symmetric, e.g. 
$g_a^\j\sqrt{1+q_a^2}\sqrt{1-\vec{V}_a^2}\bm{\phi}^\j_b(x_a)
=g_b^\j\sqrt{1+q_b^2}\sqrt{1-\vec{V}_b^2}\bm{\phi}^\j_a(x_b).$ Thus, combining individual subwall Lagrangians $\hat{L}_a$ into one\footnote{Each pairwise interaction contributes once.} (and keeping terms up to quadratic in velocities and electric charges):
\begin{align} \label{Eq:Lagrangian1}
\hat{L} = \frac12 U^{ab} (\vec{V}_a \cdot \vec{V}_b - q_a q_b) 
+ q_a W^{ab} (\vec{V}_b) ,
\end{align}
with implicit summation over the repeated indices $a$ and $b$ and
\begin{align}\label{Eq:Uaa}
U^{aa}&= \sum_{\j=1}^n g_a^\j\left(\bar{Q}^\j x_a+\bar{M}^\j+ \frac12\sum_b   g^\j_b|x_a-x_b|\right),\\
U^{ab}&=- \frac12\sum_{\j=1}^n g_a^\j g_b^\j|x_a-x_b|, \qquad \text{for\ } a\neq b,
\end{align}
and 
\begin{align}
W^{aa}&= \sum_{\j=1}^n g_a^\j
\left(\bar{Q}^\j \frac{\theta_a d\varphi-\varphi_a d\theta}{2} + r_\varphi \bar{\chi}_\theta^\j  d\theta +  r_\theta \bar{\chi}_\varphi^\j d\varphi
+  \frac12\sum_b   g_b^\j\eta_{\vec{x}_a-\vec{x}_b}\right),\\
\label{Eq:Wab}
W^{ab}&= -  \frac12\sum_{\j=1}^n g_a^\j g_b^\j\eta_{\vec{x}_a-\vec{x}_b},\qquad \text{for\ } a\neq b.
\end{align}

\subsection{Subwall Positions and Charges}
\subsubsection{Positions}
As we discussed, the motion of the subwalls is highly choreographed, since the subwalls' positions are dictated by the plane arrangement. Via Eqs.~\eqref{Eq:SpineVer}:
\begin{align}\label{Eq:ModR}
x_a=\sum_{\substack{f \\ V(f)\ni a}}c_a^f R_f,
\end{align}
where the sum is over the three faces $f$ that have $a$ as a their vertex and 
 the coefficients are $c_a^f=n_{f'}-n_{f''}$ as in Sec.~\ref{Sec:SpineV}.
In fact, the $\theta$-position of the wall is determined by the same relation
\begin{align}\label{Eq:ModTheta}
\theta_a=\sum_{\substack{f \\ V(f)\ni a}} c_a^f \Theta_f.
\end{align}

As mentioned in Sec.~\ref{Sec:Spectral}, one can consider another spectral curve $\mathbb{S}_\theta$.  Analysis of its branch points leads to the same $\varphi$ subwall position relation
\begin{align}\label{Eq:ModPhi}
\varphi_a=\sum_{\substack{f \\ V(f)\ni a}} c_a^f \Phi_f.
\end{align}

Next, we focus on understanding the relations between the electric charges $q_a$ of the subwalls.  Namely, we shall now demonstrate that they also satisfy the same relation
\begin{align}
q_a=\sum_{\substack{f \\ V(f)\ni a}} c_a^f Q_f,
\end{align}
for independent variables $Q_f$, one for each internal spine face.

\subsubsection{Electric Charges}
Let $V$ denote the set of spine vertices, $E$ -- the set of spine edges, and $F$ -- the set of spine faces. 
We shall orient the edges rightwards (and up, if vertical). For an edge $e\in E$ let $h(e)\in V$ denote its head and let $t(e)\in V$ denote its tail. 
Then the crystal vertex position $(x_a,y_a,z_a)$ is determined from the system of equations \eqref{Eq:Planes}
\begin{align}\label{Eq:xyzPlanes}
z_a&=m_f x_a+n_f y_a+R_f,
\end{align}
satisfied for all faces $f\in F$ for which $a\in V$  is a vertex of $f$: $a\in V(f).$

Taking the difference of adjacent faces, one gets the spine vertex position $(x_a,y_a)$ from the equations 
\begin{align}
&&&&M_e&:=m_{h(e)}-m_{t(e)},\nonumber\\
\label{Eq:xyEdges}
M_e x_a+N_e y_a+L_e&=0,&&\text{where}&
N_e&:=n_{h(e)}-n_{t(e)},\\
&&&&L_e&:=R_{h(e)}-R_{t(e)},\nonumber
\end{align}
for any edge $e\in E$ beginning or ending at the vertex $a\in V.$

Note that solutions $((x_a,y_a))_{a\in V}$ of \eqref{Eq:xyEdges} are in one-to-one correspondence with solutions $((x_a,y_a,z_a))_{a\in V}$ of \eqref{Eq:xyzPlanes}.  Also, Eqs.~\eqref{Eq:xyEdges} describes a system $(V_3,E_2)$ of 
\begin{itemize}
\item
$V$ points $[x_a,y_a,1]$ in $\mathbb{R}P^2$ and 
\item
$E$ lines $\{[x,y,1] | M_e x+N_e y+L_e=0 \}$ in the same in $\mathbb{R}P^2$, such that 
\item 
each point has three lines passing through it (corresponding to three edges $e$ for which $a$ is a vertex) and
\item
each line has two points on it (corresponding to the two ends of $e\in E$).
\end{itemize}
The reciprocal view of the dual $\mathbb{R}P^2$ with coordinates $[M,N,L]$ gives the $(E_2,V_3)$ system of $E$ points $[M_e,N_e,L_e]$ and $V$ lines $\{ [M,N,L] | x_a M + y_a N + L=0 \}$ such that each point has two lines through it and each line has three points on it.

Note, that the whole system is completely determined by the set of distinct points $(x_a,y_a)$, since (using the first $(V_3,E_2)$ configuration) each line is determined by two points on it.

Consider triplets $(x_a,y_a,w_a)$ with $w_a$ the (coincident) eigenvalue of $\bm{a}_0$ at the wall $a$ where this eigenvalue has a kink.  As earlier, we define $W_e:=w_{h(e)}-w_{t(e)}$ for each spine edge $e.$ Then, comparing the electric flux change across the subwall (LHS below) with the electric charge  $q_ag_a^\j$ (RHS below), one has
\begin{align}
\frac{W_{e^{out}}}{X_{e^{out}}} - \frac{W_{e^{in}}}{X_{e^{in}}}
=q_a\left(\frac{Y_{e^{out}}}{X_{e^{out}}} - \frac{Y_{e^{in}}}{X_{e^{in}}}\right).
\end{align}
This was our very definition of the electric charge $q_a g_a^\j$.  This relation implies that $p_a:=\frac{W_e-q_a Y_e}{X_e}$ is the same for any edge $e$ beginning or ending at $a$.

This implies that $u_a:=x_a p_a+y_a q_a -w_a=x_b p_a + y_b q_a -w_b$ for any edge $\overline{ab}\in E.$  Which in turn is equivalent to 
\begin{align}\label{Eq:Ghost}
P_e x_a+Q_e y_a-U_e=0,
\end{align}
for any $a\in V$ and any $e\in E$ beginning or ending at $a$.

Note, that \eqref{Eq:Ghost} is also a $(V_3,E_2)$ system.  In fact, since it has the same set of points $(x_a,y_a)_{a\in V}$ it must be the same system of projective lines and points.  Thus,
\begin{align}
M_e Q_e&=N_e P_e,&
Q_e L_e&=- U_e N_e.
\end{align}
We take the last equation as determining $U_e=-\frac{L_e}{N_e} Q_e$. The first equation, on the other hand, reads
\begin{align}\label{Eq:Str}
S_e:=-M_e q_a+N_e p_a=-M_e q_b+N_e p_b,
\end{align}
for any edge $\overline{ab}\in E.$  Since, $M_e=m_{f_{left}}-m_{f_{right}}$ and $N_e=n_{f_{left}}-n_{f_{right}}$, we have $\sum_{e, h(e)=a} S_e-\sum_{e,t(e)=a} S_e=0$ and therefore the function $\{S_e\}_e$ on edges  is potential on the dual graph, in other words, there is a function $Q_f$ such that $S_e=Q_{f_{left}}-Q_{f_{right}}$.  Here $f_{left}$ denotes the spine face to the left of the oriented spine edge $e$, and $f_{right}$ denotes the one to its right.  Substituting this into Eq.~\eqref{Eq:Str}, 
\begin{align}
m_{f_{left}}q_a + n_{f_{left}}(-p_a) + Q_{f_{left}}=m_{f_{right}}q_a + n_{f_{right}}(-p_a) + Q_{f_{right}}=:r_a.
\end{align}

We conclude that the set of triples $(q_a,-p_a,r_a)$ satisfies exactly the same system of equations as the triples $(x_a,y_a,w_a)$ with the role of $R_f$ played by $Q_f.$  Thus, $q_a$ are expressed via \eqref{Eq:SpineVer}:
\begin{align}
 q_a&=\delta n_{23}Q_1+\delta n_{31}Q_2+\delta n_{12}Q_3
 =-n_1\delta Q_{23}-n_2\delta Q_{31}-n_3\delta Q_{12}\nonumber\\
 &=\delta n_{23}\delta Q_{12}-\delta n_{12}\delta Q_{23},
\end{align}
and $q_a=\sum_{f, V(f)\ni a} c_a^f Q_f.$

\subsection{The Asymptotic Metric on the Moduli Space}\label{Sec:Metric}
Now we are ready to read off the asymptotic metric on the moduli space within each maximal cone of the secondary fan.  So far we can conclude that the effective Lagrangian \eqref{Eq:Lagrangian1}, expressed in terms of the moduli $\vec{X}_f=\left(\begin{smallmatrix}R_{f}\\ \Theta_{f}\\ \Phi_{f}\end{smallmatrix}\right)$
and independent charges $Q_f$, is
\begin{align}
\hat{L}=\frac12 c_a^f U^{ab}c_b^{f'}(\dot{\vec{X}}_f\cdot\dot{\vec{X}}_{f'} - Q_f Q_{f'})
+ Q_f c_a^f W^{ab}(\dot{\vec{X}}_{f'})c_b^{f'}.
\end{align}
To lighten our notation from here on we denote by $cUc=[(cUc)^{ff'}]$ the matrix with entries $(cUc)^{ff'}:=c_a^f U^{ab}c_b^{f'}$ and similarly for $cWc.$

The conserved charges $Q_f$ should be viewed as momenta associated with the electromagnetic phase moduli $T_f.$ In order to express the Lagrangian in terms of the moduli we perform the Legendre transform in $Q_f:$
\begin{align}
\dot{T}_f&=\frac{\partial\hat{L}}{\partial Q_f}=-(cUc)^{ff'}Q_{f'} + (cWc)^{ff'}(\dot{\vec{X}}_{f'}),\\
L&=\hat{L}-Q_{f'}\frac{\partial\hat{L}}{\partial Q_{f'}}.
\end{align}
This yields the effective Lagrangian: 
\begin{multline}
L=\frac12 (cUc)^{ff'}\dot{\vec{X}}_f\cdot\dot{\vec{X}}_{f'}\\
+\frac12\Big(\dot{T}_f - (cWc)^{f\check{f}}(\dot{\vec{X}}_{\check{f}})\Big) (cUc)^{-1}_{ff'}  \Big(\dot{T}_{f'} - (cWc)^{f'\hat{f}}(\dot{\vec{X}}_{\hat{f}})\Big),
\end{multline}
which describes free motion of a point on a manifold with the Pedersen-Poon \cite{PP88} type metric
\begin{align}\label{Eq:AsymMetric}
g=(cUc)^{ff'}d{\vec{X}}_f\cdot d{\vec{X}}_{f'}
+\Big(d{T}_f - (cWc)^{f}\Big) (cUc)^{-1}_{ff'}  \Big(d{T}_{f'} - (cWc)^{f'}\Big).
\end{align}
This is the asymptotic metric on the moduli space of the monowall.  Its terms are written in terms of the $U$ and $W$ of Eqs.~(\ref{Eq:Uaa}-\ref{Eq:Wab}) and the coefficients $c_a^f$ appearing in Eq.~\eqref{Eq:caf}.  Let us emphasize that each generic ray in the moduli space lies in a cone labelled by a regular triangulation of the Newton polygon. Thus, the end of the moduli space is divided into sectors, each with the corresponding asymptotic metric \eqref{Eq:AsymMetric}. The triangulation determines both the coefficients $c_a^f$ and the order of the subwalls' positions $x_a.$

\section{The K\"ahler Potential and the Generalized Legendre Transform}\label{Sec:Kahler}
Consider approaching the infinity of the moduli space within some maximal cone of the secondary fan.  Such a cone is specified by a triangulation Triang$(\N)$ of the Newton polygon $\N.$   
As we now demonstrate, the K\"ahler potential $K$ of the asymptotic metric is encoded in a single function:
\begin{multline}\label{Eq:GLTfn}
G(\bar{M}^\j;x_1,...,x_n) = \sum\limits_{\j=1}^h  \Bigg[ \sum\limits_{a\in\mathrm{Triang}(\N)} g_{a}^\j \left( \bar{M}^\j  \frac{x_a^2}{2} + \bar{Q}^\j  \frac{x_a^3}{6} \right) \\
 +  \frac12 \sum\limits_{\substack{a,b \\ a>b}}   g_a^\j g_b^\j \frac{(x_a-x_b)^3}{6} \Bigg].
\end{multline}
The relation is via the Generalized Legendre Transform of \cite{Lindstrom:1987ks,Hitchin:1986ea} as follows. 

Number the subwalls from left to right, so that $x_1<x_2<\ldots<x_N,$ and introduce a Laurent polynomial in the auxiliary variable $\zeta$ for each subwall 
\begin{align}
\hat{\eta}_a(\zeta) := \frac{\theta_a +\ii \varphi_a}{2\zeta}+ x_a-\frac{\theta_a -\ii\varphi_a}{2}\zeta,
\end{align}
and let 
\begin{align}
\hat{\mathcal{V}}^\j(\zeta) := (r_\theta \bar{\chi}_{\varphi}^\j - \ii r_\varphi \bar{\chi}_\theta^\j)\frac{1}{\zeta} +   \bar{M}^\j  - (r_\theta \bar{\chi}_{\varphi}^\j + \ii r_\varphi \bar{\chi}_\theta^\j) \zeta.
\end{align}
Note, that thanks to (\ref{Eq:ModR} -- \ref{Eq:ModPhi}) the polynomial coefficients $x_a,\theta_a,\varphi_a$ associated to the positions of each wall are functions of the respective moduli (and parameters) $R_{f},\Theta_{f},\Phi_{f}$:
\begin{align}
x_a&=c_a^f R_f,&
\theta_a&=c_a^f \Theta_f,&
\varphi_a&=c_a^f \Phi_f.&
\end{align}

Consider the Generalized Legendre Transform of the following auxiliary function
$$F(R_f,\Theta_f,\Phi_f)=\frac{-1}{2 \pi \ii} \oint_0 \frac{d\zeta}{\zeta} {G}\left(\hat{\mathcal{V}}; \hat{\eta}_1,...,\hat{\eta}_N\right),$$
of the parameters and of  three quarters of the moduli. Half of the complex moduli are $Z_f:=\frac{\Theta_f+\ii\Phi_f}{2}.$ The contour integration above is over a counterclockwise oriented small circle around zero. The remaining half of the  moduli  $U_f$  are related to the above coordinates by 
  \begin{align}\label{Eq:LegMom}
U_f+\bar{U}_f:=\frac{\partial F}{\partial R_f}=F_{R_f}.
\end{align}
Importantly, $F$ constructed this way is guaranteed to satisfy the Laplace type system of equations $(\partial_{Z_f}\partial_{\bar{Z}_{f'}}+\partial_{R_f}\partial_{R_{f'}})F=0.$ 

The K\"ahler potential $K$ is  the  Legendre transform of $F:$
\begin{align}
K(Z_f,U_f)=F-\sum_{f\in\mathrm{Int}(\N)} R_f (U_f+\bar{U}_f),
\end{align}
with $R_f$ on the right-hand side understood as functions of $Z_f$ and $U_f$ determined by $\eqref{Eq:LegMom}.$ As usual for the Legendre transform $K_{U_f}=-R_f$ and $K_{Z_f}=F_{Z_f}.$  This gives 
$K_{U_f \bar{U}_{f'}}=-[F_{RR}]^{-1}_{ff'}$, which is the negative inverse of the matrix $F_{RR}=(F_{R_f R_{f'}}).$ 
Also $K_{U_f\bar{Z}_{f'}}=[F_{RR}]^{-1}_{f\hat{f}}F_{R_{\hat{f}}\bar{Z}_{f'}},$ as well as
$K_{Z_f\bar{Z}_{f'}}=-(F_{R_fR_{f'}}+F_{Z_fR_{\hat{f}}}[F_{RR}]^{-1}_{\hat{f}\check{f}} F_{R_{\check{f}}\bar{Z}_{f'}}).$ 
 The resulting metric $$g^{GLT}=4(K_{Z_f\bar{Z}_{f'}}dZ_f d\bar{Z}_{f'}+K_{Z_f\bar{U}_{f'}}dZ_f d\bar{U}_{f'}+K_{U_f\bar{Z}_{f'}}dU_f d\bar{Z}_{f'}+K_{U_f\bar{U}_{f'}}dU_f d\bar{U}_{f'}),$$ 
is directly expressed in terms of $F$: 
\begin{align*}
g^{GLT}= - 4dZ_f F_{R_f R_{f'}}d\bar{Z}_{f'} 
- 4(dU_f - dZ_{\hat{f}}F_{Z_{\hat{f}}R_f})[F_{RR}]^{-1}_{ff'}
(d\bar{U}_{f'} - F_{R_{f'}\bar{Z}_{\check{f}}} d\bar{Z}_{\check{f}}),
\end{align*}
which in terms of the real moduli $R_f, \Theta_f, \Phi_f$ and $T_f:=2\mathrm{Im}\, U_f$ reads
\begin{multline}\label{Eq:GTLm}
g^{GLT}= - F_{R_f,R_{f'}}(dR_f dR_{f'}+d\Theta_f d\Theta_{f'}+d\Phi_f d\Phi_{f'})\\
    - (dT_f - W^f)
                  [F_{RR}]^{-1}_{ff'}
                 (dT_{f'} - W^{f'}).
\end{multline}
with the one-form $W^f=-\ii dZ_{\hat{f}}F_{Z_{\hat{f}}R_f}+\ii d\bar{Z}_{\hat{f}}F_{\bar{Z}_{\hat{f}}R_f}.$  The exact metric coefficients can be easily evaluated observing that 
\begin{multline}
G_{\hat{\eta}_a\hat{\eta}_b}=\sum_\j\Bigg[\delta_{ab} g_a^\j(\hat{\mathcal{V}}^\j+\bar{Q}^\j\hat{\eta}_a+
\frac12 \sum_{c, a>c} g_c^\j(\hat{\eta}_a-\hat{\eta}_c)- \frac12 \sum_{c, c>a} g_c^\j(\hat{\eta}_a-\hat{\eta}_c))\\
- \frac12 g_a^\j g_b^\j \mathrm{sign}(a-b)(\hat{\eta}_a-\hat{\eta}_b)\Bigg],
\end{multline}
and by direct calculation
\begin{multline}\label{Eq:Fff}
F_{R_f R_{f'}}=\sum_{a,b}c_a^f c_b^{f'}\frac{(-1)}{2\pi\ii}\oint\frac{d\zeta}{\zeta}  G_{\hat{\eta}_a\hat{\eta}_b}\\
  = - \sum_a c_a^f c_a^{f'} g_a^\j\left( \bar{M}^\j+\bar{Q}^j x_a+\sum_c\frac12 g_c^\j |x_a-x_c| \right)
     + \sum_{a,b}\frac12c_a^f g_a^\j c_b^{f'}  g_b^\j|x_a-x_b|,
\end{multline}
\begin{multline}
F_{R_f Z_{f'}}=\sum_{a,b}c_a^f c_b^{f'}\frac{(-1)}{2\pi\ii}\oint\frac{d\zeta}{\zeta}  \frac{1}{\zeta} G_{\hat{\eta}_a\hat{\eta}_b}\\
  = \sum_a c_a^f c_a^{f'}g_a^\j\Big(r_\theta\bar{\chi}_\varphi+\ii r_\varphi\bar{\chi}_\theta + \bar{Q}^\j \frac{\theta_a-\ii\varphi_a}{2}
 +\frac12\sum_c g_c^\j\frac{\theta_{ac}-\ii\varphi_{ac}}{2}\mathrm{sign}(x_a-x_c)\Big)\\
-\sum_{a,b}\frac12 c_a^f g_a^\j c_b^{f'}g_b^\j\frac{\theta_{ab}-\ii\varphi_{ab}}{2} \mathrm{sign}(x_a-x_b).
\end{multline}
Using
$F_{R_f\bar{Z}_{f'}}=\sum_{a,b}c_a^f c_b^{f'}\frac{1}{2\pi\ii}\oint\frac{d\zeta}{\zeta}  (-\zeta)G_{\hat{\eta}_a\hat{\eta}_b}=\overline{F_{R_f Z_{f'}}},$ 
one has
\begin{multline}\label{Eq:Wf}
W^f=2 \mathrm{Im}\, dZ_{\hat{f}}F_{Z_{\hat{f}}R_f}=\sum_{a}c_a^f g_a^\j\left( r_\theta\bar{\chi}_\varphi d\varphi_a+r_\varphi\bar{\chi}_\theta d\theta_a+\bar{Q}^\j\frac{\theta_a d\varphi_a-\varphi_a d\theta_a}{2}\right.\\
+\frac12\sum_c g_c^\j\frac{\theta_{ac} d\varphi_a-\varphi_{ac} d\theta_a}{2}\mathrm{sign}(x_a-x_c)\\
\left.-\sum_{b}\frac12g_b^\j \frac{\theta_{ab} d\varphi_b-\varphi_{ab} d\theta_b}{2}\mathrm{sign}(x_a-x_b)\right),
\end{multline}
which exactly matches Eqs.~(\ref{Eq:Uaa}-\ref{Eq:Wab}).

Thus, we directly verified that the resulting GLT metric \eqref{Eq:GTLm} with \eqref{Eq:Fff} and \eqref{Eq:Wf} exactly  matches the asymptotic metric  \eqref{Eq:AsymMetric} obtained from the subwall dynamics:
\begin{align}
g^{GLT}= g.
\end{align}

\section{From  the GLT Function to the Cut Volume}\label{Sec:Relation}

\subsection{Cut Volume}
We make use of the Lawrence formula \cite{Lawrence} for the volume of a simple convex polytope $P=\{ x\in\mathbb{R}^n\, |\, \vec{a}_i\cdot\vec{x}\leq b_i, i=1,\ldots,m \}:$
\begin{align}\label{Eq:Lawrence}
\mathrm{Vol}(P)=\sum_{\vec{v}\in\mathrm{Vert}(P)} N_{\vec{v}},
\end{align}
which is a sum of signed volumes of simplices with
\begin{align}\label{Eq:simplex}
N_{\vec{v}}=\frac{1}{n!}\frac{(\vec{c}\cdot\vec{v}+d)^n}{\gamma_1\gamma_2\ldots\gamma_n|\det(a_{i_1},a_{i_2},\ldots,a_{i_n})|},
\end{align}
the signed volume of the simplex with its apex at $\vec{v}$ and its base in the base plane $\vec{c}\cdot\vec{x}+d=0.$ 
Here 
\begin{itemize}
\item
$\vec{v}$ is one of the vertices of $P$ with exactly $n$ of the planes $\vec{a}_{i_1}\cdot\vec{x}=b_{i_1},\ldots,\vec{a}_{i_n}\cdot\vec{x}=b_{i_n}$ passing through it, 
\item
the corresponding simplex is cut out by these $n$ planes and the base plane $\vec{c}\cdot\vec{v}+d\geq0$ with some fixed vector $\vec{c}$ not normal to any of the polygon planes,
\item
the constants $\gamma_1,\ldots,\gamma_n$ are the coefficients in the decomposition 
$$\vec{c}=\gamma_1 \vec{a}_{i_1}+\ldots+\gamma_n \vec{a}_{i_n}.$$
\end{itemize}

Let us gain some appreciation of this formula \eqref{Eq:simplex} by proving it.  In dimension $n=3$, let $\vec{e}_1,\vec{e}_2,\vec{e}_3$ be the simplex edges emanating from its main vertex $\vec{v}$ and ending on its base plane  $\vec{c}\cdot\vec{x}+d=0.$ 
Let $\vec{b}$ be a point in this base plane and let $\vec{v}_0=\vec{v}-\vec{b}$ be its height, i.e. $\vec{c}\cdot\vec{b}+d=0$ and $\vec{c}\cdot\vec{v}+d=\vec{c}\cdot\vec{v}_0.$ 
Clearly the symplex volume is
\begin{align}
\mathrm{Vol}_{\vec{v}}=\frac{1}{3!}\det(\vec{e}_1,\vec{e}_2,\vec{e}_3).
\end{align}

The corresponding vectors $\vec{a}_1,\vec{a}_2,\vec{a}_3$ are normal to respective simplex faces, and thus each $a_i$ is proportional to the vector product $\vec{e}_j\times \vec{e}_k$ of the two edges of that simplex face:
\begin{align}
(\vec{a}_1,\vec{a}_2,\vec{a}_3)=(\vec{e}_2\times\vec{e}_3,\vec{e}_3\times\vec{e}_1,\vec{e}_1\times\vec{e}_2)
\left(\begin{smallmatrix}
\alpha_1&0&0\\
0&\alpha_2&0\\
0&0&\alpha_3
\end{smallmatrix}\right).
\end{align}
To lighten our notation let $Det=\det(\vec{e}_1,\vec{e}_2,\vec{e}_3).$ By construction 
$\vec{c}=(\vec{a}_1,\vec{a}_2,\vec{a}_3)\gamma$, thus
\begin{align}
\begin{pmatrix}
\gamma_1\\ \gamma_2\\ \gamma_3
\end{pmatrix}=(\vec{a}_1,\vec{a}_2,\vec{a}_3)^{-1}\vec{c}=\left(\begin{smallmatrix}
\alpha_1^{-1}&0&0\\
0&\alpha_2^{-1}&0\\
0&0&\alpha_3^{-1}
\end{smallmatrix}\right)\frac{1}{Det}
\begin{pmatrix}
\vec{e}_1^{\,T}\\
\vec{e}_2^{\,T}\\
\vec{e}_3^{\,T}
\end{pmatrix}\vec{c},
\end{align}
giving $\gamma_j=\frac{\vec{e}_j\cdot\vec{c}}{\alpha_j Det}.$ 
Noting that 
$(\vec{e}_2\times\vec{e}_3,\vec{e}_3\times\vec{e}_1,\vec{e}_1\times\vec{e}_2)
={Det}\cdot\begin{pmatrix}
\vec{e}_1^{\,T}\\
\vec{e}_2^{\,T}\\
\vec{e}_3^{\,T}
\end{pmatrix}^{-1},$ we have the Lawrence formula take the form
\begin{multline}
N_{\vec{v}}=\frac{1}{3!}\frac{(\vec{c}\cdot\vec{v}+d)^3}{\gamma_1\gamma_2\gamma_3|\det(\vec{a}_{1},\vec{a}_{2},\vec{a}_{3})|}
=\frac{1}{3!}\frac{(\vec{c}\cdot\vec{v}_0)^3}{\frac{\vec{c}\cdot\vec{e}_1}{\alpha_1 Det}\frac{\vec{c}\cdot\vec{e}_2}{\alpha_2 Det}
\frac{\vec{c}\cdot\vec{e}_3}{\alpha_3 Det}
|\alpha_1\alpha_2\alpha_3|\frac{|Det|^3}{|\det(\vec{e}_{1},\vec{e}_{2},\vec{e}_{3})|}}\\
=\pm\frac{1}{3!} Det=\pm\mathrm{Vol}_{\vec{v}}.
\end{multline}
We used $\vec{c}\cdot\vec{v}_0=\vec{c}\cdot\vec{e}_1=\vec{c}\cdot\vec{e}_2=\vec{c}\cdot\vec{e}_3,$ since $\vec{c}$ is normal to the base plane continaining $\vec{e}_i-\vec{e}_j$ and $\vec{v}_0-\vec{e}_j.$ 

The signs in the Lawrence formula are chosen already so that the individual simplex volumes contribute with different signs 
and the polygon volume does not depend on the choice of the base plane $\vec{c}\cdot\vec{x}+d=0.$

Let us choose a high horizontal plane $z=M$ for a very large value  $M.$ The cut volume is the difference of the volume of the (convex) regularized blocked crystal
\begin{align}
\bar{\mathcal{C}}_0=\{(x,y,z)\,|\, m_fx+n_fy+R_f\leq z\leq M, f\in\mathrm{Int}(\N)\},
\end{align} 
and the (convex) regularized cut crystal
\begin{align}
\bar{\mathcal{C}}_{cut}=\{(x,y,z)\,|\, m_fx+n_fy+R_f\leq z\leq M, f\in\N\}.
\end{align}
The Lawrence formula applies to both $\bar{\mathcal{C}}_0$ and $\bar{\mathcal{C}}_{cut}$ and thus the cut volume  is 
\begin{align}
\Vol=\mathrm{Vol}(\bar{\mathcal{C}}_{0})-\mathrm{Vol}(\bar{\mathcal{C}}_{cut})
=\sum_{\substack{(ppp)\\(ppt)}}N_{\vec{v}}-\sum_{\substack{(ppi)\\(pii)\\(iii)\\(ppt)}}N_{\vec{v}}
=\sum_{\substack{(ppp)}}N_{\vec{v}}-\sum_{\substack{(ppi)\\(pii)\\(iii)}}N_{\vec{v}},
\end{align}
where the first sum is over the vertices $(ppp)$ at which three perimeter planes (i.e. planes corresponding to the points in Per$(\N)$) meet or $(ppt)$ at which two perimeter and one top plane meet.  The last sum is over the vertices $(**i)$ involving an internal plane as well as the vertices $(ppt)$ involving the top plane.  The latter $(ppt)$ contributions cancel (as, indeed, the cut volume does not depend on the choice of the high top plane). The remaining $(ppp)$ contributions are moduli independent, thus, up to a constant, the volume we are interested in is 
\begin{align}
\bar{\Vol}=-\sum_{a\in V}N_{\vec{v}_a},
\end{align}
with $\vec{v}_a$ the apex of the cone cut out by $m_fx+n_fy-z\leq -R_f$ for three internal points $(m_f,n_f)\in\N$ forming the $\Delta_{a}$ triangle of the triangulation.   In the Lawrence formula we choose $\vec{c}=(1,0,0)^T$ and $d=0$, and have $(a_{f_1},a_{f_2},a_{f_3})=\left(\begin{smallmatrix} m_1&m_2&m_3\\n_1&n_2&n_3\\-1&-1&-1\end{smallmatrix}\right).$ 
Thus, $\det(a_{f_1},a_{f_2},a_{f_3})=m_{31}n_{21}-m_{21}n_{31}=-\delta_{123}$, where $\delta_{123}$ is the area of the triangle\footnote{Here we use our conventions of the footnote on page \pageref{FT:AreaNorm}, i.e. $\delta_{123}$ is twice the conventional triangle area.} $\left((m_1,n_1),(m_2,n_2),(m_3,n_3)\right).$  And the relevant factors read off from $(\gamma_1,\gamma_2,\gamma_3)^T=(a_{f_1},a_{f_2},a_{f_3})^{-1}(1,0,0)^T$ are 
$\gamma_1=n_{23}/\delta_{123}, \gamma_2=n_{31}/\delta_{123}, \gamma_3=n_{12}/\delta_{123}$.  The resulting volume formula is 
\begin{align}\label{Eq:VolF}
\bar{\Vol}=-\sum_{a\in V}N_{\vec{v}_a}=-\sum_{a\in V} \frac{\delta_{a}^2}{n_{12}n_{23}n_{31}} \frac{x_a^3}{6},
\end{align}
with the triangle $((m_i,n_i))_{i=1,2,3}$ positively oriented.  

\subsection{GLT Function}
The GLT function \eqref{Eq:GLTfn} is
\begin{multline}
\begin{aligned} 
G&=\frac{1}{12}\sum_\j\left(\sum_a g_a^\j(6\bar{M}^\j x_a^2+2\bar{Q}^\j x_a^3)+\sum_{a>b}   g_a^\j g_b^\j(x_a-x_b)^3\right)\\
& =\frac{1}{12}\sum_\j\left(\sum_a 3 g_a^\j x_a^2(2\bar{M}^\j + 2 \bar{Q}^\j x_a)\right.
\end{aligned} \\
 +3\sum_{a>b}   g_a^\j g_b^\j(x_a^3-x_a^2x_b)
 +3\sum_{b>a}   g_a^\j g_b^\j(x_b x_a^2-x_a^3)\\
\left. -2\sum_a  g_a^\j x_a^3(2\bar{Q}^\j+\sum_{b | b<a}   g_b^\j - \sum_{b | b>a}   g_b^\j)\right).
\end{multline}
In terms of the Higgs field \eqref{Eq:AllStat}, this reads 
\begin{align}\label{Eq:IntG}
G=\sum_a\sum_\j g_a^\j\bm{\phi}^\j(x_a)\frac{x_a^2}{2} 
 - \sum_\j\sum_a g_a^\j\frac{x_a^3}{6}\left(2\bar{Q}^\j+  \sum_{b | b<a} g_b^\j -   \sum_{b | b>a} g_b^\j\right).
\end{align}
For any given subwall $a$ all $U(1)$ factors $\j$ which have a nonzero charge $g_a^\j$ have the same value $\bm{\phi}^\j(x_a)$, while the charges themselves satisfy $\sum_\j g_a^\j=0$, since\footnote{We suppose for concreteness that $n_2>n_1>n_3.$} $n_{13}$ of the $U(1)$ factors have $g_a^\j=\frac{m_{13}}{n_{13}}-\frac{m_{23}}{n_{23}}$ and $n_{21}$ of the $U(1)$ factors have $g_a^\j=\frac{m_{21}}{n_{21}}-\frac{m_{23}}{n_{23}}:$
\begin{align}
\pm\sum_\j g_a^\j &=n_{13} \left(\frac{m_{13}}{n_{13}}-\frac{m_{23}}{n_{23}} \right)+n_{21} \left(\frac{m_{21}}{n_{21}}-\frac{m_{23}}{n_{23}} \right) \\ \nonumber
&=m_{13}+m_{21}-m_{23}=0.
\end{align}
Thus, the first term in \eqref{Eq:IntG} vanishes.  If we let $S_a^\j$ denote the magnetic flux in the $\j$-th $U(1)$ factor  to the right of the $a$-th subwall, then $g_a^\j=S_a^\j-S_{a-1}^\j$ and  $\bar{Q}^\j=(S_0^\j+S_N^\j)/2$, as defined in \eqref{Eq:Brd}.  Thus, the second term in \eqref{Eq:IntG} is a telescoping series: $\sum_{b | b<a}g_a^\j=S_{a-1}^\j-S_0^\j$ and 
$\sum_{b | b>a}g_a^\j=S_{N}^\j-S_a^\j$, therefore,  the last term becomes $  \sum_{\j,a} g_a^\j\frac{x_a^3}{6} \left( S_a^\j+S_{a-1}^\j \right)=\sum_{\j,a}   \frac{x_a^3}{6}((S_a^\j)^2-(S_{a-1}^\j)^2).$  Summing over the $U(1)$ factors, 
$\sum_\j \left((S_a^\j)^2-(S_{a-1}^\j)^2 \right) = n_{21}\left(\frac{m_{21}}{n_{21}}\right)^2+n_{32}\left( \frac{m_{32}}{n_{32}}\right)^2 - n_{31}\left(\frac{m_{31}}{n_{31}}\right)^2=\frac{(m_{21}n_{32}-n_{21}m_{32})^2}{n_{21}n_{32}n_{31}}.$  As a result 
\begin{align}
G=-\sum_a   \frac{x_a^3}{6}\frac{\delta_{a}^2}{n_{12}n_{23}n_{31}}.
\end{align}
Here $\delta_a$ is twice the conventional area of the triangle $(m_i,n_i), i=1,2,3$ associated with $a$-th subwall.

Comparing to \eqref{Eq:VolF} we conclude that the GLT function is equal to the cut volume:
\begin{align}
G=  \Vol.
\end{align}

\section{Outlook}\label{Sec:Conclusions and Outlook}
The effective dynamics of a monopole wall are given by the  electromagnetic interaction of its constituents.  In the low speed approximation it produces the effective Lagrangian from which we read off the resulting asymptotic moduli space metric.  We proved that the K\"ahler potential of this metric is the Generalized Legendre Transform of the regularized crystal volume cut out by the plane arrangement.  The latter volume can be easily read off from the monopole charges, parameters, and moduli.

The remaining challenge is to find the K\"ahler potential for the whole moduli space.  With this goal in mind we now pose some questions and take the liberty of  making some speculations.

There is a more refined volume function at hand that could capture some of the K\"ahler potential subleading asymptotic behavior.  Consider the Ronkin function
\begin{align}
\mathcal{R}^\varphi_{C_{m,n}}(x,y)=\frac{1}{(2\pi\ii)^2}\oint\limits_{\substack{|s|=\exp{\frac{x}{r_\theta}}\\|t|=\exp{\frac{y}{r_\theta}}}}\ln|P(s,t)| \frac{ds}{s}\frac{dt}{t}.
\end{align}
It is linear outside of the amoeba $\mathcal{A}:=\{(\ln|s|,\ln|t|) : P(s,t)=0 \}$ with $\mathcal{R}^\varphi_{C_{m,n}}(x,y)=mx+ny+\tilde{R}_{m,n}.$  Note that as moduli approach infinity  $\tilde{R}_{m,n}\rightarrow R_{m,n}.$ These planes lead to a function $\tilde{m}(x,y):=\max_{(m,n)\in\N}\{mx+ny+\tilde{R}_{m,n}\}.$  
The region above the graph of $\mathcal{R}^\varphi_{C_{m,n}}$ is the {\em melted crystal.} 
One can consider the volume of the region $\{(x,y,z) : \tilde{m}(x,y)<z<\mathcal{R}^\varphi_{C_{m,n}}(x,y)\}$ and use this melted volume $\Vol_{melt}$ instead of the cut volume $\Vol$ used in this paper.  For large moduli these two volumes $\Vol_{melt}$ and $\Vol$ are exponentially close to each other and thus produce the same asymptotic.

One might seek to combine the two Ronkin functions $\mathcal{R}^\varphi$ and $\mathcal{R}^\theta$, for example, incorporating both $\theta$ and $\varphi$ spectral curves $\mathbb{S}^\theta$ and $\mathbb{S}^\varphi$ to encode the complete K\"ahler potential.

The relation between the two Legendre transforms that we used can be summarized in the following diagram:
\begin{center}
\begin{tabular}{ccc}
Tent function over $\N$ & $\xleftrightarrow{\text{Legendre Transform}}$ & 
\begin{tabular}{c}Cut Crystal\\ Surface $z=M(x,y)$\end{tabular}\\
$\displaystyle\left\updownarrow\vphantom{\int_A^B}\right.$? & & $\displaystyle\left\updownarrow\vphantom{\int_A^B}\right.$ \\
\begin{tabular}{c} K\"ahler Potential\\ $K(Z_f,U_f)$\end{tabular} & $\xleftrightarrow[\text{Transform}]{\text{Generalized Legendre}}$ & Cut Volume $\Vol(R_f)$
\end{tabular}
\end{center}
This leads to a question: Is there a more direct relation between the tent function and the K\"ahler potential?  Is there a natural physical meaning of the Legendre transform of the Ronkin function in this context?

Let us conclude with a conjecture for the auxiliary function $G$ for the exact K\"ahler potential.  To begin, we define the {\em Twistor Spectral curve} $S^{\mathrm{tw}}$ \cite{Cherkis:2007xs} via the Hitchin scattering problem \cite{Hitchin:1982gh}.  The space of oriented lines in the covering space $\mathbb{R}^3$ of the base space $\mathbb{R}\times S^1\times S^1$ is the minitwistor space $T\mathbb{P}^1$.   Each line $\ell$ is determined by the unit vector of its direction $\hat{n}$, (which determines the point with the complex coordinate $\zeta$ on the Riemann sphere $\mathbb{P}^1$) and the line's displacement from the origin (which is a point in the tangent plane at $\hat{n}$ with coordinate $\eta\in T_\zeta\mathbb{P}^1=\mathbb{C}\cup\{\infty\}$).  For each line, consider the scattering problem $(D_{\hat{n}}+\Phi)\psi=0.$ For some lines this problem has an $L^2$ solutions.  These lines are called the {\em spectral lines}. Each line in $\mathbb{R}^3$ is a point in $T\mathbb{P}^1$ and the set of all spectral lines forms a curve $S_0^{\mathrm{tw}}$ in $T\mathbb{P}^1.$  
Since our initial problem is invariant under discrete shifts in the $\theta$ and $\varphi$ directions, the curve $S_0^{\mathrm{tw}}$ descends to a curve $S^{\mathrm{tw}}$ in the quotient space $\mathcal{Z}:=T\mathbb{P}^1/2\pi( r_\theta \hat{n}_\theta\mathbb{Z}\oplus r_\varphi \hat{n}_\varphi\mathbb{Z}),$
which is the space of geodesics in $\mathbb{R}\times S^1\times S^1.$ 
Let $\{\eta_1(\zeta),\ldots,\eta_n(\zeta)\}$ be the local branches of this twistor spectral curve.  We conjecture that 
\begin{align}
G=\Vol_{melt}\left(\frac{\eta_1}{\zeta},\ldots,\frac{\eta_n}{\zeta}\right)
\end{align}
produces the exact K\"ahler potential. 

The challenge in using such a relation is that even for the conventional monopoles in $\mathbb{R}^3$ the twistor curve is notoriously difficult to find, as it should satisfy a complicated `triviality condition'.  In addition, for monowalls, the curve $S^{\mathrm{tw}}$ is contained in the minitwistor space $\mathcal{Z}$ that is non-Hausdorff, while its cover $S^{\mathrm{tw}}_0\subset T\mathbb{P}^1$ is of infinite genus.  Some recent approaches, such as in \cite{2019arXiv190203551M}, provide promising perspectives on this problem.

\section*{Acknowledgments}
SCh is grateful to the organizers of the 2019 workshop ``Microlocal Methods in Analysis and Geometry'' at CIRM--Luminy and to the Institute des Hautes \'Etudes Scientifiques, Bures-sur-Yvette where the final stages of this work were completed. 
SCh received funding from the  European Research Council under the European Union Horizon 2020 Framework Programme (h2020) 
through the ERC Starting Grant QUASIFT (QUantum Algebraic Structures In Field Theories) nr. 677368. RC thanks the Marshall Foundation for her Dissertation Fellowship funding.

\bibliographystyle{amsalphaurl}
\bibliography{Kahler.bib}

\end{document}